\documentclass[11pt]{article}

\newcommand{\figref}[1]{Fig.\ref{#1}}

\newcommand{\tit}[1]{\textit{#1}} % Italic Font
\newcommand{\tbf}[1]{\textbf{#1}} % Bold Font

\newcommand{\ie}{\textit{i.e }}

\newcommand{\mcal}[1]{\mathcal{#1}}
\newcommand{\mbf}[1]{\mathbf{#1}}
\newcommand{\mbb}[1]{\mathbb{#1}}
\newcommand{\bsym}[1]{\boldsymbol{#1}}

\newcommand{\mathleft}{\@fleqntrue\@mathmargin0pt}
\newcommand{\mathcenter}{\@fleqnfalse}

\newcommand{\Cd}{\nabla} % 3-dim Cov-Deriv
\newcommand{\cd}{\mbf{D}} % 2-dim Cov-Deriv
\newcommand{\Ld}[1]{\mcal{L}_{\bsym{#1}}} % 2-dim Cov-Deriv
\newcommand{\pd}[1]{\partial_{#1}} % 2-dim Cov-Deriv

\newcommand{\ko}{\mathring{k}} % Traceless part of the tangential projection of the extrinsic curvature K_{ab}

\newcommand{\normal}{\hat{n}}
\newcommand{\ndot}{\dot{\normal}}
\newcommand{\lapse}{\hat{\alpha}} % The lapse of the 2+1 foliation
\newcommand{\shift}{\hat{\beta}} % The shift of the 2+1 foliation
\newcommand{\JT}{J^{(\perp)}} % Normal part of the current vector
\newcommand{\Jp}{J^{(||)}} % Tangential part of the current vector

\usepackage[english]{babel}
\usepackage{graphicx}
\usepackage{xcolor}
\usepackage{color,amssymb,color}
\usepackage[bookmarks=true,bookmarksopen=true,colorlinks=true,breaklinks=true,linkcolor=blue,citecolor=blue]{hyperref}
\usepackage{amsmath,amsfonts,amssymb,mathrsfs,amsthm}
\usepackage[affil-it,max2]{authblk}
\usepackage{enumerate}   
\usepackage{cleveref}
\usepackage{caption}
\usepackage{subcaption}
\usepackage[export]{adjustbox}
\usepackage{pifont}
\DeclareMathOperator{\dist}{dist}

\usepackage[margin=2.7cm]{geometry}
\usepackage{cite}

\newtheoremstyle{break}% name
  {}%         Space above, empty = `usual value'
  {}%         Space below
  {\itshape}% Body font
  {}%         Indent amount (empty = no indent, \parindent = para indent)
  {\bfseries}% Thm head font
  {.}%        Punctuation after thm head
  {\newline}% Space after thm head: \newline = linebreak
  {}%         Thm head spec

\newtheoremstyle{inline}
  {}{}                 % above/below space
  {\itshape}           % body font (upright text)
  {}                   % indent
  {\bfseries}          % head font
  {.}                  % punctuation after head
  { }                  % space after head (just a space, no newline)
  {}                   % head spec

\theoremstyle{break}
\newtheorem{theorem}{Theorem}[section]

\newtheorem{prop}{Proposition}[section]
\newtheorem{pcoro}{Corollary}[prop]
\newtheorem{definition}{Definition}[section]
\newtheorem*{definition*}{Definition}

\theoremstyle{inline}

\title{Numerical stability of the Hyperbolic Formulation of the Constraint Equations for $\mathbb{T}^3$ cosmological spacetimes}

\author[1]{Alejandro Estrada-Llesta\footnote{email: alejandro.estrada.llesta@univie.ac.at}}
\author[2]{Cristhian Martinez-Duarte \footnote{email: cisthian.duarte@correounivalle.edu.co}}
\author[3]{Leon Escobar-Diaz\footnote{email: leon.escobar@correounivalle.edu.co}}
\affil[1]{Department of Mathematics, Universität Wien, Austria.}
\affil[2]{Department of Physics, Universidad Del Valle, Colombia.}
\affil[3]{Department of Mathematics, Universidad Del Valle, Colombia.}
\begin{document}
\maketitle

\begin{abstract}
% In this work, we numerically construct initial data sets for cosmological spacetimes with a spatial topology of $\mathbb{T}^3$, which are generally inhomogeneous. 
% To do so, we implement a pseudo-spectral approach based on the discrete Fourier transform for numerically solving Einstein's constraint equations in an algebraic-hyperbolic form. 
% We explore the advantages and disadvantages of this method by comparing the numerical solutions with known analytical initial data sets. 
% Additionally, we perform an stability analysis of the system to gain deeper understanding of the problem. 
% Finally, we numerically obtain new families of initial data sets through manipulation of the original system by imposing restrictions in some of the variables.\\

%(sometido antes )
% In this work, we study the viability of the algebraic-hyperbolic formulation of the Einstein's constraint equations to construct initial data sets for inhomogeneous cosmological spacetimes with $\mathbb{T}^3$ topology. 
% To do so, we implement a pseudo-spectral approach based on the discrete Fourier transform for numerically and explore the advantages and disadvantages of this method by comparing the numerical solutions with known analytical initial data sets.
% Additionally, we perform an stability analysis of the system to gain deeper understanding on the limitations of the proposed scheme. 
% Finally, we numerically obtain new families of initial data sets through manipulation of the original system by imposing restrictions on some variables.
%%%

In this work, we study the algebraic–hyperbolic formulation of the Einstein constraint equations for the numerical construction of initial data sets for inhomogeneous cosmological spacetimes with $\mathbb{T}^3$ topology.
We implement a pseudo-spectral method of lines based on the discrete Fourier transform, and we find that the scheme exhibits pathological instabilities.
Through linear stability analysis, we prove that the instabilities are unavoidable for any spacetime sufficiently close to FLRW. In contrast, we find that this approach can be stable for Gowdy spacetimes depending on the initial time choice. 
Additionally, we present evidence that certain subclasses of the algebraic–hyperbolic formulation, when combined with a Fourier-based method of lines, are numerically stable, thus offering a potential new path for computing initial data sets for inhomogeneous cosmological spacetimes.
\end{abstract}

\section*{Introduction }

The evolution of cosmological spacetime is mathematically formulated as an initial value problem within the framework of general relativity. This requires solving two coupled sets of equations: the evolution equations and the constraint equations. The latter determine the \textit{initial data} for the evolution problem. Solving these constraint equations, either analytically or numerically, is a challenging task, as they generally constitute a system of coupled, nonlinear partial differential equations.\\

A standard numerical approach to this problem is the Lichnerowicz–York conformal method, which employs a conformal transformation to recast the constraint equations as a set of coupled, nonlinear elliptic partial differential equations (for comprehensive discussions, see \cite{alcubierre2008introduction,Baumgarte}).
This method has proven to be highly effective for constructing initial data sets in a wide range of asymptotically flat spacetime configurations, such as binary black hole systems.
Its success relies largely on the simplifying assumption that the initial spatial hypersurface is conformally flat, which facilitates the prescription of suitable boundary conditions at spatial infinity.
In spherically symmetric cases, this assumption is exact, while in more general situations it remains a useful approximation that yields physically consistent and numerically stable initial data.\\

However, when extending this approach to cosmological spacetimes, several conceptual and technical difficulties arise. Cosmological models typically possess compact or otherwise nontrivial spatial topologies and are not asymptotically flat, meaning that the notion of spatial infinity—and hence the associated boundary conditions—no longer applies. As a result, the conformal decomposition central to the Lichnerowicz–York method becomes significantly less constrained: there is no canonical or physically motivated choice for the conformal metric. Consequently, the elliptic equations derived from the conformal transformation depend sensitively on the selected conformal background, while the global geometry of the cosmological slice imposes additional consistency conditions absent in the asymptotically flat case. These features render the direct application of the Lichnerowicz–York framework to cosmological initial data considerably more intricate.\\

To overcome these challenges, alternative formulations have been proposed. One possibility is to impose periodic boundary conditions, as in models with toroidal spatial topology, thereby replacing asymptotic flatness with spatial periodicity. Such approaches enable the numerical construction of cosmological initial data that remain consistent with the global structure of a compact universe. In this context, Garfinkle and Mead \cite{Garfinkle2020} applied the conformal method in a cosmological setting with periodic boundary conditions, demonstrating that this strategy can successfully reproduce spatially closed universes within the conformal framework. Although this approach represents a valuable step toward constructing cosmological initial data sets compatible with periodic spatial topologies, it remains limited to rather special cases of the Einstein constraint equations and requires restrictive assumptions on both the matter content and the choice of free conformal data.\\ 

Recent works have sought to overcome some of these limitations by extending or modifying the conformal approach. 
For instance, Aurrekoetxea, Clough, and Lim \cite{Aurrekoetxea2022} proposed the CTTK method, which replaces the elliptic Hamiltonian constraint with an algebraic relation for the mean curvature, improving numerical stability in inhomogeneous cosmological settings. A broader perspective on these developments can be found in \cite{aurrekoetxea2025cosmology}, which surveys current efforts to generalize conformal and constraint-based formulations beyond the asymptotically flat regime and toward more realistic cosmological configurations.\\ 
 
Another promising approach to solve the constraint equations was introduced by Rácz in a series of works \cite{racz2014cauchy,racz2014bianchi,racz2015constraints}. In contrast to the conformal method, this formulation recasts the equations into parabolic–hyperbolic or algebraic–hyperbolic systems.
One of the main advantages of this approach is that it requires freely specifiable data only on a two-dimensional surface, thereby avoiding the need to solve globally coupled elliptic equations.
In compact-topology cosmologies, where spatial infinity is absent, elliptic formulations are computationally expensive and poorly suited to periodic boundary conditions.
By contrast, the hyperbolic character of Rácz’s formulation enables localized evolution and a natural implementation of fully periodic spectral schemes, offering a potentially more efficient and flexible framework for constructing consistent initial data sets.\\
 
The primary motivation of this work is to explore the applicability of the Algebraic–Hyperbolic Formulation (AHF) of the Einstein constraint equations in cosmological settings, particularly in compact topologies where traditional elliptic approaches may encounter conceptual or numerical limitations. Specifically, we apply the AHF to cosmological models with $\mathbb{T}^3$ spatial topology, focusing on two physically relevant cases: $\mathbb{T}^3$–Gowdy spacetimes and perturbed FLRW universes. Both represent benchmark configurations in modern numerical relativity, serving as testbeds to investigate nonlinear gravitational dynamics and relativistic backreaction in compact cosmologies.\\

$\mathbb{T}^3$–Gowdy spacetimes, admitting two commuting Killing vectors, provide an exact framework for studying nonlinear gravitational wave dynamics and singularity formation in spatially compact universes. They exhibit rich nonlinear phenomena such as spike formation and curvature blow-up near singularities \cite{Amorim2009, Gambini2005, ringstrom2010cosmic}. In contrast, perturbed FLRW models describe globally homogeneous and isotropic backgrounds with small-scale inhomogeneities that drive structure formation. Within the $\Lambda$CDM paradigm, such perturbations give rise to the observed cosmic microwave background anisotropies and the large-scale distribution of matter \cite{Ma_Berts}. Adopting a $\mathbb{T}^3$ topology enables a fully periodic numerical domain, allowing a consistent treatment of nonlinear effects beyond perturbation theory. Recent developments in numerical cosmology \cite{Macpherson2019, Bentivegna2024} have emphasized the importance of compact, boundary-free formulations for exploring fully relativistic inhomogeneous universes.\\

Within this context, we numerically implement the AHF (which we briefly introduce in Section~\ref{section:AHF_presentation}). 
We use the Method of Lines (MoL) based on Fourier differentiation, which we describe in Section~\ref{section:numerical_approach}.
This implementation aims to provide a tool to obtain accurate solutions to the resulting system of partial differential equations. 
In Section~\ref{section:test_error}, we see that although the method is able to reproduce $\mathbb{T}^3$–Gowdy and perturbed FLRW (PFLRW) solutions, it presents some worrisome convergence issues for the perturbed FLRW case.
For this reason, in Section~\ref{sec:Stability}, we study the linear stability of the MoL approach applied to this system.
There (see Thrm.~\ref{teorema}), we prove that the ill behavior observed for PFLRW is not coincidental but unavoidable. 
This result comes directly from the spectral structure of the Fourier differentiation matrices and the nature of the ODE integration methods (in particular, Runge-Kutta). \\

Even though this is a very pessimistic result, \tit{it is not possible to find initial data for PFLRW spacetimes with the approach presented in Sections~\ref{section:AHF_presentation} and \ref{section:numerical_approach}}, it holds under three sufficient conditions: (1) using Fourier-based MoL, (2) applying it to the AHF PDE system and (3) restricting the problem to PFLRW spacetimes.
Therefore, it might be possible to obtain satisfactory constraint solutions by modifying any of these three pieces. 
That is what we partially address in the remainder of the work. 
In Section~\ref{sec:GowdyStability}, we provide strong numerical evidence of the stability and viability of the method if the reference spacetime is changed, or the system obtained from the AHF is modified by imposing additional conditions, as we do in Section~\ref{section:new_id}.
Finally, in Section~\ref{section:discussion}, we summarize our main findings and discuss their implications for future applications of the AHF in cosmological contexts.

\section{Algebraic-hyperbolic formulation of the constraints}\label{section:AHF_presentation}

Let us consider a smooth $3$-dimensional manifold with the spatial topology of $\mathbb{T}^3$, which we identify by $\Sigma$, endowed with a Riemannian metric $\gamma_{ab}$, and second fundamental form $K_{ab}$ with respect to some Lorentzian smooth  $4$-dimensional manifold $M$; i.e., $\Sigma$ will be embedded in $M$. The tuple $(\gamma_{ab}, K_{ab}, \rho, J_a)$ represents an initial data for the evolution of Einstein equations   on the manifold $M$ if the following tensorial equations on $\Sigma$ are satisfied\footnote{We use Latin indices $a,b,c...$ to denote tensor components of $3$-dimensional manifolds taking values of $1,2,3$. On the other hand, we use $i,j,k,...$, running for $1,2$ to denote tensor components of 2d-manifolds.}:  
\begin{eqnarray}
      \label{eq:HC_operator}
    R  + K^2 - K_{ab}K^{ab} -16 \pi \rho &=& 0  ,\\
  \label{eq:MC_operator}
    \Cd_bK^b_{~a} - \Cd_a K -8 \pi J_a  &=& 0,
\end{eqnarray}
where $\gamma^{ab}$ is the inverse of $\gamma_{ab}$, $K := \gamma^{ab} K_{ab} $ is the mean curvature of $\Sigma$ with respect  to $M$, $R $ is the intrinsic curvature of $\Sigma$, and $\nabla_ a$  is the covariant derivative operator compatible with $\gamma_{ab}$. The quantities $\rho$ and $J_a$ are the energy and the current densities respectively. The tensor equations (\ref{eq:HC_operator}) and (\ref{eq:MC_operator})  are commonly referred to as the \textit{Hamiltonian} and \textit{Momentum} constraints equations respectively \cite{alcubierre2008introduction}.\\

Following the $2+1$ decomposition of  Racz's work \cite{racz2015constraints,racz2014bianchi}, we can transform the above equations into an algebraic-hyperbolic system as follows. We assume that $\Sigma  \simeq \mathbb{T}^3= \mathbb{S}^1 \times \mathbb{T}^2$, and that it admits a complete foliation of surfaces $S_{r} \simeq \mathbb{T}^2$  parameterized by level surfaces of a smooth, positive and monotone increasing function $ r: \Sigma \to \mathbb{R}^+ $, i.e., we choose the foliation $S_{r}$ such that 

\begin{equation*}  
\Sigma = \bigcup\limits_{r=0}^{r_f}  S_r ,
\end{equation*}
with $S_{r_i} \cap S_{r_j} = \emptyset$ for $i \neq j$. 
The vector $\hat n^a$ will represent the unitary  normal vector to the surfaces $S_r$. In analogy with the standard $3+1$ decomposition of the spacetime (see \cite{Baumgarte}), we choose adopted to the foliation coordinates $(r,x^1, x^2)$ such that $r^{a}$ is the tangent vector to the curves generated by the parameter $r$ and that satisfies the relation $r^a \nabla_a r:= 1$. Further, we will assume that these coordinates cover $\Sigma$ completely. \\

From the above, it follows that the unitary normal vector to each surface $S_r $ can be expressed as 
\begin{equation*}\label{eqc:1}
\hat n^{a} := \hat \alpha^{-1}  \left (r^{a} - \hat \beta^{a} \right),
\end{equation*}
where $\hat \alpha$ and $\hat \beta^a $ will be respectively called the ``\textit{lapse function}'' and  the ``\textit{shift vector}'' associated to the vector $ r^a $ respect to the surfaces $S_{r}$. From the general theory of hypersurfaces (see for instance \cite{do1992riemannian}), it is clear that the projector operator
\begin{equation}\label{ec:proyector_equation}
h^{a}_{\ b} := \delta^{a} _ {\ b} - \hat n^{a} \hat n_{b},
\end{equation}
with $\delta^{a} _ {\ b}$ being the standard Kronecker delta, induces a  metric on $S_r$ as (see \cite{Wald})
\begin{equation}\label{eqc:2}
h_{ab} = \gamma_{ab} -  \hat n_a   \hat n_b,
\end{equation}
and a covariant derivative $\cd_a$ compatible with the metric $h_{ab}$: 
$$\cd_a := h_{a}^{\ b} \nabla_b.$$ 

Next, we decompose the second fundamental form  $K_{ab}$  in terms of   $\hat \alpha $, $\hat \beta^a $ and $h_{ab} $ as follows
\begin{equation}\label{ec:mean_curvature_decomposition}
K_{ab} = Z \hat n_{a} \hat n_{b} + \hat n_{a}   Y_{b} +  \hat n_{b} Y_{a} + \left( \mathring{ k } _{ab} + \dfrac{1}{2} h_{ab} X \right),
\end{equation}
where
\begin{equation}\label{ecs:formulae_of_decomposition}
Z := \hat n^{a} \hat n^{b} K_{ab}, \quad Y_{a} := h^{b}_{\ a } \hat n^{c} K_{bc}, \quad  
\mathring{ k } _{ab} + \dfrac{1}{2} h_{ab} X := h^{c}_{\ a} h^{d}_{\ b} K_{cd}, 
\end{equation} 
with $ h^{ab} \mathring{ k } _{ab}=0$. Replacing (\ref{ec:mean_curvature_decomposition}) into (\ref{eq:HC_operator}) and (\ref{eq:MC_operator}), and after some computations, we can express the constraint equations in terms of the geometric quantities $Z, X, Y_a$ in adapted coordinate frame $(\partial_r, \partial_{x^1}, \partial_{x^2})$  as  
\begin{align}
  \label{drXeq}
  \pd{r}X & = \Ld{\shift}X + \lapse \left( \cd_jY^j - 2Y^j\ndot_j + (Z - \frac{1}{2}X)H_j^{~j} - H_{ji}\ko^{ji} - 8 \pi \JT \right),\\
  \label{drYeq}
  \pd{r}Y_i &= \Ld{\shift}Y_i + \lapse \left( \frac{1}{2}\cd_iX + \cd_iZ - Y_i H^j_{~j} - Z \ndot_i + \frac{1}{2}\ndot_iX + \ndot^j \ko_{ij} - \cd^j\ko_{ij} + 8 \pi \Jp_i \right),\\
  \label{Zeq}
  Z  &= \frac{1}{2X}\left( 2Y_iY^i -\frac{1}{2}X^2 + \ko_{ij} \ko^{ij} - R + 16 \pi \rho \right).
\end{align}
Here $i,j,=1,2$ are indices related to the coordinates $(x^1, x^2)$ on the surfaces $S_{r}$. From now on we will refer to this system simply as the \textit{hyperbolic constraints}. Note that since $Y_{a}$ is totally tangential to $S_{r}$, $Y_{r}=0$. In this equations we have defined $\JT   := \hat n^{a} J_{a}$, $\Jp_i := h^{a}_{\ i} J_{a}$, $\dot{\hat{n}}_a := \hat n^b \cd_b \hat  n_a = - \cd_a (\text{ln} \hat \alpha)$,
  $\mathscr{L}_{\bsym{\hat{\beta}}}$ denotes the Lie derivative along the shift vector $\hat{\beta}^i$, and 
$H_{ij}$ is the second fundamental form of the $2-$dimensional surfaces $S_{r}$ with respect to $\Sigma$. This tensor and its trace are given in terms of $h_{ij}$ and $\hat{n}^{a}$ as, respectively, 
\begin{equation*}\label{ec:segundaformaenspheres}
H_{ij} =  \dfrac{1}{2 \hat\alpha} \left( \partial_r h_{ij} - 2 \cd_{(i} \hat n_{j)} \right),  \quad H:= h^{ij} H_{ij} .
\end{equation*}
Following \cite{racz2015constraints, racz2014bianchi}, it can be proved that given the fields $h_{ij}$, $\hat{\alpha}$, $\hat{\beta}i$, $\ko^{ij}$, $\rho$ and $J_i$ in $\Sigma$, the hyperbolic system comprises a first order hyperbolic system of PDEs in the variables $X$ and $ Y_{i}$ if the condition 
\begin{equation}\label{ec:hyperboliccondition}
   Z X < 0
\end{equation}
holds for all $r$ in some interval $[r_0,r)$.
In other words, Eq.\eqref{ec:hyperboliccondition} is a sufficient (although not necessary) condition for the analytical stability of the AHF of the constraint equations as an "evolution" problem along the variable $r$.
The local existence and uniqueness of its solution is guaranteed in $[r_0,r)$ for some initial data of $X$ and $ Y_{i}$ at $r_0$. \\

Summarizing: First, we freely chose the fields $\gamma_{ij}$, $\ko_{ij}$, $\rho$ and $J_i$. Second, we use the $2+1$ decomposition to obtain $h_{ij}$, $\hat{\alpha}$ and $\hat{\beta}_i$. In particular, in the \tit{adapted-to-foliation} coordinated frame, the components of $\gamma_{ab}$ can can be written explicitly as
\begin{equation}\label{gamma3d}
  \gamma_{ab} =  \begin{pmatrix}
  \hat \alpha + \hat \beta_{m} \hat \beta^{m} & \hat \beta_{i} \\ 
  \hat \beta_{j} & h_{ij}
\end{pmatrix}. 
\end{equation}
Third, we use these field for  solving  the hyperbolic constraints to obtain the field $X$ and $Y_{i}$ up to some value  $r_f$. For doing this, however, we have to  specify the initial conditions for $X$ and $Y_i$ at the initial value of $r_0$. This add another freedom to the solution of the system. 
Finally, we reconstruct the extrinsic curvature $K_{ab}$ by means of  (\ref{eqc:2}) and (\ref{ec:mean_curvature_decomposition}), which, in adapted-to-foliation coordinates, we can write as 
\begin{equation}\label{K3d}
 K_{ab} = \begin{pmatrix}
 \hat \beta^l \hat \beta^m k_{lm} + 2 \hat{\alpha} \hat \beta^l Y_l + \hat{\alpha}^2 Z &  \hat \beta^l k_{li} + \hat{\alpha} Y_i \\
   \hat \beta^l k_{li} + \hat{\alpha} Y_i & k_{ij}
\end{pmatrix}.
\end{equation}\\

As a result, we obtain the tensor components of $K_{ab}$ such that the initial data set $(\gamma_{ab},K_{ab},\rho,J_{a})$ satisfy the  Hamiltonian (\ref{eq:HC_operator}) and Momentum (\ref{eq:MC_operator}) constraints in the coordinated region of $[r_0,r_f) \times \mathbb{T}^{2}$.  Note that in principle $[r_0,r)$ does not necessarily cover all the domain $\mathbb{S}^1$, hence, it is not guaranteed that for any initial data of the field $X$ and $Y_i$ we can find solutions of the system for all $r\in\mathbb{S}^1$. In other words; the theorem only guarantees the local existence and uniqueness of the solutions, therefore, more study of this system is required in order to understand the necessary conditions for obtaining initial data sets in the all domain $\Sigma$. 
With that aim, in this work, we conduct a numerical exploration of the hyperbolic constraints by considering different choices of the free fields $h_{ij}$, $\hat{\alpha}$, $\hat{\beta}_i$, $\mathring{k}_{ij}$, $\rho$, $J_i$ and some initial values for  $X$ and $Y_i$. In the next section, we will briefly introduce our numerical infrastructure for solving the hyperbolic constraints with the appropriate boundary conditions of $\Sigma \simeq \mathbb{S}^1 \times \mathbb{T}^{2}$.

\section{Numerical approach}
\label{section:numerical_approach}

We will make two assumptions based on the topology we want to describe: First, the adapted coordinates $(r,x_1,x_2)$ on $\Sigma \simeq \mathbb{S}^1 \times \mathbb{T}^2$ are global. 
The notation choice $(r,x_1,x_2)$ is made for consistency with the literature and the previous section, it does not represent spherical coordinates. 
Second, all fields $u:= u(r,x_1,x_2)$ on any $S_r$, for some fixed $r$, are periodic along the coordinates $(x_1,x_2)$.  
For simplicity, from now on we will refer to $(x_1,x_2)$ as the \textit{angular coordinates}, and $r$ as the \textit{radial} or foliation coordinate. Under these two assumptions, we can implement the pseudo-spectral Fourier method to solve the system of PDE equations (\ref{drXeq})-(\ref{Zeq}), as we will briefly describe in the rest of the section. 
Additionally, as we will display in Section\ref{sec:ErrorFunctions}, the $3$-dimensional constraint equations, equations \eqref{eq:HC_operator} and \eqref{eq:MC_operator}, will be used as solution error measurement. Since these evaluations requires the computation of radial derivatives,  
we will assume that the fields $u(r,x_1,x_2)$ are also periodic along the $r$ variable so we can use Fourier differentiation to compute $\partial_ru$.

\subsection{Angular discretization and derivatives} 
\label{sec:FourierDifferentiation}

Let us start by defining the angular grid. We will assume that the domain for the coordinates $x_1$ and $x_2$ is some closed region given by $[-L,L] \times [-L,L]$, which we denote as $[-L,L]^2$. Thus, all fields defined at $S_r$ are $2L$-periodic along each angular coordinate. Next, we define a grid on $[-L,L]^2$ as the set of tuples $\{(x_{1i},x_{2j})\}_{i,j=0}^{N_{x_1},N_{x_2}}$, where $N_{x_1}$ and $N_{x_2}$ are fixed positive integers. Assuming that the nodes $x_{1\,i}$ and $x_{2\,j}$ are equally spaced, we can explicitly write them as $x_{1\,i} = -L + ih_{1}$ and $x_{2\,j} = -L + jh_{2}$, where $h_{1} = 2L/N_{x_1}$ and $h_{2} = 2L/N_{x_2}$ are the step sizes in the directions of the variables $x_{1}$ and $x_{2}$ respectively.\\

We will denote the values of the fields $u$ on the angular grid as $u(r)_{ij} := u(r,x_{1\,i},x_{2\,j})$, assuming $r$ fixed, as  
 \begin{align}\label{eq:2d_fields}
u(r)_{ij} &:= \frac{1}{(2L)^2}\sum_{ k_1 = -N_{x_1}/2+1  }^{N_{x_1}/2} \left( \sum_{k_2 = -N_{x_2}/2+1}^{N_{x_2}/2}   \tilde{u}(r)_{k_1k_2} \ b_{k_1,k_2}(x_{1},x_{2})  \right) \Bigg|_{x_1 = x_{1i},~ x_2 = x_{2i}} ,
\end{align}
where we have used $b_{k_1,k_2}(x_{1},x_{2})$ to denote the $2$-dimensional Fourier basis
\begin{equation*}
b_{k_1,k_2}(x_{1},x_{2}) := e^{\mathrm{i} \frac{\pi}{L}(k_1 x_{1} + k_2 x_{2} ) },
\end{equation*}
with $\mathrm{i}$ being the imaginary unit and  
$\tilde{u}(r)_{k_1k_2}$ the Fourier spectral coefficients associated with $u(r)_{ij}$ (which can be obtained using the discrete Fourier transform \cite{Canuto}). Thus, the derivative of $u(r)_{ij}$ with respect the angular coordinates $\{x_s\}_{s=1,\,2} = \{x_1,x_2\}$ (hereon \textit{angular derivatives}) can be easily obtained by differentiating eq. \eqref{eq:2d_fields} before evaluation
\begin{align*}
\label{eq:FourierDifferentiation}
\partial_{x_s} u(r)_{ij}  = \frac{1}{(2L)^2}\sum_{ k_1 = -N_{x_1}/2+1  }^{N_{x_1}/2} \left( \sum_{k_2 = -N_{x_2}/2+1}^{N_{x_2}/2}  \mathrm{i}\frac{\pi}{L} k_n \ \tilde{u}(r)_{k_1k_2} \ b_{k_1,k_2}(x_{1},x_{2}) \right)\Bigg|_{x_1 = x_{1i},~ x_2 = x_{2i}}.
\end{align*}
This process is known as Fourier differentiation. See \cite{Kopriva} for a detailed discussion of this subject.
 
\subsection{Radial discretization}\label{sec:evolutionscheme}

The derivatives along the $r-$coordinate during the solution process will be computed using a standard initial value problem ODE methods. 
For the implementation tests on Section~\ref{section:test_error}, we use $4$th order Runge-Kutta method (RK4). However, as we will observe, this method presents convergence issues which are the main motivation for the stability analysis of Section~\ref{sec:Stability}.
Although we use explicit methods for the evolution of the non-linear system \eqref{drXeq}-\eqref{Zeq}, we also consider the relevance of implicit ODE schemes on the stability analysis of the method.
Here, we describe the general setup for single-step methods that we will work with.
\\

Let $\bsym{U}(r,x_1,x_2)=\left(X(r,x_1,x_2),Y_1(r,x_1,x_2),Y_2(r,x_1,x_2)\right)$ be the solution vector for the system (\ref{drXeq})-(\ref{Zeq}). 
We denote by
\begin{equation*}
   \bsym{U}^{(n)}_{ij}:= \left(X(r_n,x_{1i},x_{2j}),Y_1(r_n,x_{1i},x_{2j}),Y_2(r_n,x_{1i},x_{2j})\right) ,
\end{equation*}
the numerical approximation of the solution vector at $(x_{1i},x_{2j})\in S_{r_n}$ and we use 
\begin{equation}\label{eq:SchematicODESystem}
  \partial_r\bsym{U}_{ij} = \bsym{F }\left( r, \bsym{U}_{ij} , \partial \bsym{U}_{ij} \right),
\end{equation}
as schematic representation of the system (\ref{drXeq})-(\ref{Zeq}) when the angular discretization has already being carried out. 
In this sense, the symbol $\partial$ in eq. \eqref{eq:SchematicODESystem} represents the Fourier differentiation described in Section~\ref{sec:FourierDifferentiation} applied to each component of $\bsym{U}$.\\

Then, we can compute the vector solution at   $(x_{1i},x_{2j})\in S_{r_{n+1}}$ by
\begin{equation}\label{eq:DiscreteODESystem}
  \bsym{U}^{ (n+1) }_{ij} = \bsym{f}\left( r_{n}, \bsym{U}^{(n)}_{ij} , \partial \bsym{U}^{(n)}_{ij}; r_{n+1}, \bsym{U}^{(n+1)}_{ij} , \partial \bsym{U}^{(n+1)}_{ij}; h \right),
\end{equation}
where $h$ is the step size of the method, $\partial \bsym{U}^{(n)}_{ij}$ are the spatial derivatives of $\bsym{U}^{(n)}_{ij}$ at $r = r_n$, and $\bsym{f}$ encodes the system of equations (\ref{drXeq})-(\ref{Zeq}) and the chosen ODE method. 
In the general, $\bsym{f}$ might depend on $\bsym{U}^{(n+1)}_{ij}$ as expressed in eq. \eqref{eq:DiscreteODESystem}. In such a case, eq. \eqref{eq:DiscreteODESystem} is said to be an implicit method.
Solving this equation for implicit methods result in a increase in the computational cost mainly for non-linear systems. 
For this reason we employ 4th order Runge-Kutta method for the numerical integration of the non-linear system of equations (\ref{drXeq})-(\ref{Zeq}). 
\\

% \AC{INTRODUCE IN PARRAGRAPH BELOW IF NEEDED: However, unlike standard evolution problems where the temporal variable can evolve indefinitely, our independent variable $r$ can only take values from $r_0$ to some $r_f$ in order to cover the entire manifold $\mathbb{S}^1$. Therefore, and in order to preserve the periodicity of the  of the solutions $X$ of $Y_i$ along $\mathbb{S}^1$, we will employ a ``forward-backward" solution that we explain as follows.}
At this point, it is important to remember that our solution vector $\bsym{U}$ take values in the $3$-dimensional torus $\mathbb{T}^3$. 
This means that $\bsym{U}$ must be periodic along each coordinate $r$, $x_1$ and $x_2$. 
This periodicity is already enforced in the angular dimensions ($x_1$ and $x_2$) by the Fourier differentiation process which is only valid for periodic functions and preserves the periodicity. 
However, standard ODE schemes do not preserve periodicity along the integration variable (in our case $r$) and it has to arise \tit{naturally} from the dynamic of the system.
In order to enforce the periodicity of the solution vector along the radial coordinate $r$, we employ a \tit{forward-backward} strategy to compute the solutions. \\

The \tit{forward-backward} strategy consists of splitting the solution of the system \eqref{eq:SchematicODESystem} in two parts: first using the ODE scheme to evolve from $r = r_0$ to $r = (r_0+r_f)/2$ in the \tit{forward} direction in $r$ (increasing $r$; positive step size), while the second part is computed from $r = r_f$ to $r = (r_0+r_f)/2$ in the \tit{backward} direction (decreasing $r$; negative step size). If the vector solution is periodic and well-behaved in $r$, the two solutions must match at the midpoint $r=(r_0+r_f)/2$ . 
This approach is intended to avoid divergent numerical solutions due to numerical instabilities. However, as we shall see in Section~\ref{section:test_error}, it will not be enough to achieve solutions in all situations. This behavior that will be clarified in Section~\ref{sec:Stability}.\\

Finally, for the evaluation of the \tit{goodness} of the solution we must verify if the $3$-dimensional constraint equations \eqref{eq:HC_operator} and \eqref{eq:MC_operator} are satisfied. 
Therefore, after evolution, the $3$-dimensional quantities must be recovered to construct the full $3$-dimensional extrinsic curvature $K_{ab}$.
And, in particular, for the evaluation of the Momentum constraint, we need to evaluate derivatives of $K_{ab}$.
To do so, as explained in Section~\ref{sec:ErrorFunctions}, we will use $3$ dimensional Fourier differentiation. 
Since an accurate evolution needs considerably more points than an accurate Fourier partial derivative evaluation, we define the number of radial-evolution steps in terms of the number of radial-derivative nodes in such a way that the radial-evolution step size $\Delta r$ is defined as
\begin{equation*}
\Delta r = \frac{2(r_f-r_0)}{F~N_r}.   
\end{equation*} 
Here, $N_r$ is the number of nodes used to evaluate partial derivatives, and $F$ is the quotient between the number of radial-evolution steps and the number of Fourier nodes in the radial direction. $F$ will also be referred to as $Factor$ throughout this work. 
This allows us to extract the values to be used during differentiation from the list of evolution results just by taking the steps $\bsym{U}^{(n)}$ with $n/F \in \mathbb{Z}$. \\ 

\subsection{Error functions}
\label{sec:ErrorFunctions} % former sec:error

We finish this section by specifying the error or discrepancy functions that  we use to measure the error of the numerically obtained initial data. 
Let us assume that we have  a numerical approximation of a solution of the hyperbolic system (\ref{drXeq})-(\ref{Zeq}) denoted by $\{ \bsym{U}(r_n,x_1,x_2)\}^{n=N_r}_{n=0}$. Additionally, assume we have chosen the radial coordinate in $r\in(-L,L)$, which allows us to have a $3$-dimensional grid on the cube $[-L,L]^3:=[-L,L]\times[-L,L]\times[-L,L]$ where our approximation is defined.\\

We want to measure \tit{how far} this approximations is from satisfying the constraint equations (eqs. (\ref{eq:HC_operator}) and (\ref{eq:MC_operator})). To do so, we will take advantage of the periodicity of the fields and we will use an approach based on the $3$-dimensional Fourier transform as follows.
First, we will denote $x_0=r$. Thus, we can denote the evaluations of the  fields $u$ on the grid as $u_{lij} := u(x_{0l},x_{1i},x_{2j})$. 
In analogy to (\ref{eq:2d_fields}) for $2$-dimensional fields, we can write the $3$-dimensional samples of the $3$-dimensional fields as
\begin{align*}
u_{lij} &:= \frac{1}{(2L)^3}
\sum_{k_0 = -N_{x_0}/2+1}^{N_{x_0}/2} 
\sum_{ k_1 = -N_{x_1}/2+1  }^{N_{x_1}/2} \sum_{k_2 = -N_{x_2}/2+1}^{N_{x_2}/2} 
\tilde{u}_{k_0k_1k_2} \ b_{k_0,k_1,k_2}(x_0,x_{1},x_{2}) \Bigg|_{(x_0,x_1,x_2)=(x_{0\,l},x_{1\,i},x_{2\,j})},
\end{align*}
where $\tilde{u}_{k_0k_1k_2}$ are the spectral coefficients and the $b_{k_0,k1,k2}(x_{0},x_{1},x_{2})$ are the 3-D Fourier basis
\begin{equation*}
b_{k_0 k_1,k_2}(x_{0},x_{1},x_{2}) := e^{\mathrm{i} \frac{\pi}{L}( k_0 x_{0} + k_1 x_{1} + k_2 x_{2} ) }.
\end{equation*}
Using this approach, we can compute all the spatial derivatives of the  $\{ \bsym{U}(r_n,x_1,x_2)\}^{n=N_r}_{n=0}$ on the grid points of $\Sigma$.  
Second, by means of eqs. (\ref{gamma3d}) and, (\ref{K3d}), we can reconstruct the components of the $3$-dimensional tensor $\gamma_{ab}$ and $K_{ab}$. Thus, if define
\begin{eqnarray}   
   \mcal{H}(\gamma_{ab},K_{ab}) &:=& R  + K^2 - K_{ab}K^{ab} -16 \pi \rho,\label{hamiltonian_error}\\
  \mcal{M}_a(\gamma_{ab},K_{ab}) &:=& \Cd_bK^b_{~a} - \Cd_a K -8 \pi J_a,\label{momentun_error}
\end{eqnarray}
we can use the Fourier differentiation of $3$-dimensionalfields to obtain evaluations of these functions $\mcal{H}(\gamma_{ab},K_{ab})_{lij}$ $\mcal{M}_c(\gamma_{ab},K_{ab})_{lij}$ at each the $3$-dimensional grid point. Therefore, we define the Hamiltonian and momentum error, respectively;
\begin{eqnarray*}    
\text{Hamiltonian error} &=& ||  H( h_{ab},K_{ab})_{lij}||, \\
\text{Momentum error} &=&||\mcal{M}_c(h_{ab},K_{ab})_{lij} || , \ c = 1,2,3, 
\end{eqnarray*}
where $|| \cdot ||$ denotes the norm of the maximum (discrete version of the infinity norm)
\begin{equation}
\label{eq:max_norm_def}
||f|| = \max_{ l,i,j }\{|f_{lij}|\}
\end{equation}
for all the values $f_{lij}$ of the function $f$ on the $3$-dimensional grid.  
Validation and discussion on our implementation of this evaluation can be found in Appendix~\ref{appendix:Constriant_Checker}.

\section{Testing the numerical approach}
\label{section:test_error}

In this section, we will explore the feasibility of the numerical approach presented in Section~\ref{section:numerical_approach} to solve the hyperbolic constraint equations (\ref{drXeq})-(\ref{Zeq}). For this purpose, we will numerically reproduce some known analytical solutions of the constraint equations and investigate the convergence of the method. 
Additionally, some comments about the aliasing error in the numerical solutions and its control with filter strategies can be found in Appendix~\ref{appendix:aliasing_and_filter}.

\subsection{Some exact solutions}\label{sec:exact_solutions}
 
% In this work we will consider the following known spacetimes metrics that we will use as test beds for our numerical approach.   
Our main cases of study are $\mbb{T}^3$ Gowdy and PFLRW spacetimes (introduced below).
Additionally, to test our implementation of the $3$-dimensional constraint equations, we also use metrics for Minkowski and Gowdy spacetimes under specific coordinate transformations. 

\begin{enumerate}
  \item \tbf{The Gowdy spacetime:} The Gowdy $\mbb{T}^3$ spacetimes are solutions of the vacuum Einstein equations that describes an expanding universe with gravitational radiation information (see for instance \cite{ringstrom2010cosmic} for a review of this spacetime). This metric is used by M. Alcubierre \tit{et. al.} in \cite{Alcubierre_Testbeds} as a test for numerical relativity codes in a strong field context. In global-periodic coordinates $(r,x_1,x_2)$ in $\mbb{T}^3$, this metric can be written as
  \begin{equation}
    \label{Gowdy_Metric}
    g_{\mu\nu} =  \left(
                  \begin{matrix}
                    -\frac{e^{\frac{1}{2}Q(t,r)}}{\sqrt{t}} & 0 & 0 & 0 \\
                    0 & \frac{e^{\frac{1}{2}Q(t,r)}}{\sqrt{t}}  & 0 & 0 \\
                    0 & 0 & t e^{-P(t,r)} & 0 \\
                    0 & 0 & 0 & t e^{P(t,r)} \\
                  \end{matrix}
                  \right).
  \end{equation}
As in \cite{Alcubierre_Testbeds}, we take $P$ and $Q$ as
\begin{equation}
       \label{PandQ_Gowdy}
       \begin{split}
         P(t,r) &= J_0( 2 \pi t) \cos(2 \pi r) \text{ and }\\
          Q(t,r) &= -2 \pi t J_0(2\pi t) J_1(2\pi t) \cos(2 \pi r)^2 +2 \pi^2 t^2 (J_0(2\pi t)^2 +\\
           &+ J_1(2\pi t)^2) - \frac{1}{2} ((2 \pi)^2 (J_0(2\pi)^2 + J_1(2\pi)^2 ) - 2 \pi J_0(2\pi) J_1(2\pi)),\\
       \end{split}
     \end{equation}
where $J_0(\cdot)$ and $J_1(\cdot)$ are the first kind Bessel functions. 
In order to keep consistency with the notation of Section~\ref{section:AHF_presentation}, we relabel Cartesian coordinates such that the foliation coordinate $r$ corresponds to the $z$ coordinate in \cite{Alcubierre_Testbeds} and $x = x_1$ and $y = x_2$. 

Since this (along with PFLRW spacetime introduced next) is the spacetime we will apply the AHF, it is important to mention whether the hyperbolicity condition Eq.\eqref{ec:hyperboliccondition} is satisfied. 
As mention in Section~\ref{section:AHF_presentation}, this condition is sufficient (but not necessary) to guarantee the analytical stability of the system.
For this particular case, the product $XZ$ depends on $t$, and as we will see in Section~\ref{sec:Stability}, Eq.\eqref{ec:hyperboliccondition} is only satisfied in all the domain for some $t$ choices. 
This will be critical for the numerical stability and is discussed in Section~\ref{sec:Stability}.

\item\tbf{The Perturbed-Friedman-Robertson-Walker spacetime (PFLRW):} This case, (as Minkowski spacetime introduced next), does not have $\mbb{T}^3$ topology. However, it is expected that, at large scales, different portions of the spacetime behave similarly in order to display cosmological homogeneity and isotropy. This approach is widely used in cosmology \cite{Peebles,Ma_Berts,Dodelson_2ed,DurrerCMB} where the Fourier transform is used to move the linear perturbation equations from the physical-coordinate space to the Fourier-wave-number space. In particular, we will work with the scalar PFLRW metric as given in \cite{Ma_Berts} or \cite{Australians2017,Australians2019}, in the Newtonian or longitudinal gauge, conformal time and periodic coordinates $(r,x_1,x_2)$ as follows
\begin{equation}
\label{SPFLRW_Metric}
g_{\mu\nu} = a(\eta)^2
              \begin{pmatrix}
                -(1+2\psi) & 0 & 0 & 0 \\
                0 & 1 - 2\phi & 0 & 0 \\
                0 & 0 & 1 - 2\phi & 0 \\
                0 & 0 & 0 & 1 - 2\phi \\
              \end{pmatrix}.
\end{equation}

The potentials $\psi$ and $\phi$ are, in general, functions of the conformal time $\eta$ and the coordinates $(r,x_1,x_2)$.
Following \cite{Australians2017}, we will consider a simple form of the potentials
$\psi = \phi = \phi_0\sum_a\sin( \pi x_a/L), \text{ with } \{x_a\}_{a = 0}^3 = \{r,x_1,x_2\}$ ($x_0 := r$). The parameter $\phi_0$ will determine the amplitude of the potentials and will be set as $10^{-8}$.

Since eq. (\ref{SPFLRW_Metric}) is a non-vacuum solution of Einstein equations, we need to set the sources $\rho\text{ and } J_i$. During the experiments of the successive sections, unless other thing stated, the sources $(\rho,J_i)$ will be computed from the constraint equations (\ref{eq:HC_operator})-(\ref{eq:MC_operator}). Once the $4-$dimensional metric is chosen, $\gamma_{ab}$ and $K_{ab}$ can be computed through the $3+1$ decomposition and used to solve the constraint equations for $\rho \text{ and } J_i$. Therefore, we obtain expressions for the sources in terms of the metric functions. In this case, for instance, $\rho \text{ and } J_i$ will be functions of the potential $\phi$. This procedure gives us analytical expressions for the source functions and allows us to close the system of Eqs. (\ref{drXeq}-\ref{Zeq}) to numerically reproduce the analytical solutions.
Finally, throughout this document, to evaluate Eq. (\ref{SPFLRW_Metric}) numerically, we set the value of the scale factor and its derivative as  $a(\eta)=a'(\eta) = 1$. 
Unfortunately, evaluation of the impact of these values on the initial data is outside the scope of this work.

Lastly, the hyperbolicity condition, Eq.\ref{ec:hyperboliccondition}, is never satisfied in this case. 
By explicitly computing the quantities $X$ and $Z$ it is possible to see that, when $\phi=0$, 
$$XZ = \frac{1}{2}X^2 = \frac{2\dot{a}^2}{a^4}\geq 0$$
for every value of $\eta$~\footnote{The final conclusion, $X\,Z\geq0$, also holds for arbitrary $\phi$, however, since we are in the perturbation regime, it is enough to see that the hyperbolicity condition is not fulfilled in this case.}.

\end{enumerate}
\subsection{Radial evolution: convergence tests and errors} \label{sec:convergence_test}

Here we perform a preliminary exploration of the scheme we have presented during the previous sections. 
We will evaluate two different error metrics: (1) evaluations of the $3$-dimensional constraint equations (as presented in Appendix~\ref{appendix:Constriant_Checker}), and (2) convergence evaluation of the radial integration method. \\

In order to evaluate the Hyperbolic Constraints, eqs. (\ref{drXeq}) and (\ref{drYeq}), we need to set all the free fields that are present in these equations ($\lapse$, $\shift_i$, $H_{ij}$, etc.). 
The explicit expressions for each of these functions can be obtained by applying the $2+1$ decomposition to the spatial metric $\gamma_{ab}$ and to the extrinsic curvature $K_{ab}$ (see Section \ref{section:AHF_presentation}). %\footnote{Although the expressions for these functions and tensors is not always too complex, since there are more than ten functions to display for each metric, it is not worth to write them here.}.
Additionally, this decomposition of a known $4-$dimensional metric into its $3+1$ and $2+1$ quantities allows us to obtain analytical solutions for $X$ and $Y_i$ that we can use to compare with our numerical solutions. 
To this end, we define the error of any $2-$dimensional quantity $u(r)$ that take values from $S_r$ as
\begin{equation*}
    E_u(r) = ||u^{(teo)}(r) - u^{(num)}(r)|| \ ,
\end{equation*}    
for each radial step, where $u^{(teo)}(r)$ is the analytical solution obtained from the decomposition of the $4-$dimensional metric, $u^{(num)}(r)$ is the numerical approximation to $u(r)$ and $||\cdot|| $ is the $2-$dimensional version of the norm of the maximum defined in eq. (\ref{eq:max_norm_def}).\\

Once we have closed the system by setting all the free fields, we can proceed to find numerical approximations. As exposed in Section~\ref{section:numerical_approach}, the Fourier differentiation method allows us to transform the PDE system formed by eqs. (\ref{drXeq}) and (\ref{drYeq}) into an ODE system that can be solved by using any numerical integration method. To evaluate if this method is converging, we perform convergence tests following chapter $9$ of \cite{alcubierre2008introduction}. 
This process consists of repeatedly solve the system  while varying the radial step size and compare the obtained solutions in the way we describe in the following paragraph.\\

For a square angular grids, we define a resolution $\Xi = \{\Delta r, N\}$ as the object that contains the information of the radial and angular grids. The numerical solution obtained by using the resolution $\Xi$ will be denoted by
\begin{equation*}
U_{\Xi}(r,x_1,x_2):=\left(X_{\Xi}(r,x_1,x_2),Y_{1,\,\Xi}(r,x_1,x_2),Y_{2,\,\Xi}(r,x_1,x_2)\right). 
\end{equation*}
We can also define a collection of resolutions $\{\Xi_i = \{\Delta r_i, N\}\}_{i}$ for a fixed number of angular nodes $N$. If $\Delta r_{i+1} = \Delta r_{i}/2$, then the theory of one-step numerical methods for ODEs (see \cite{Stoer_Bulirsch}) implies the following relation holds for solutions attained using, for example, a 4th order Runge-Kutta method
\begin{equation}
\label{Di_eq}
 D_i(r) - D_{i+1}(r) = \log_2\left(\frac{||U_{\Xi_i} - U_{\Xi_{i+1}}||}{||U_{\Xi_{i+1}} - U_{\Xi_{i+2}}||}\right) = 4, 
\end{equation}
where $D_i(r) = \log_2(||U_{\Xi_i} - U_{\Xi_{i+1}}||)$. 
This means that by increasing the resolution by a factor of $2$ in the radial direction ($2$ times more radial steps), a numerical solution must approach to the next one at the rate of $2^4$.
With the aim of making more clear the differences on eq. \eqref{Di_eq}, we can also define the quantities $C_i(r) = D_i(r) - D_{i+1}$ which, in the RK4 convergence regime, must fulfill $C_i \approx 4.$\\

For the convergence tests of this section we have used the time steps $\Delta r_i = \{2^iN\}_{i=3}^7$ where $N$ is the number of nodes per dimension in the angular grid.

% \NOTE{There is no analytic vs numeric plot. A plot $Err_{max}$ vs N will be enough to show that PFLRW anticonverges}
\begin{enumerate} 

    \item\tbf{Test for the Gowdy metric:} In this case, the functions depends only on the radial variable. Therefore, if there are no modifications in the free data, the Hyperbolic Constraints Formulation becomes a system of ODEs. 
    In this case the numerical method is convergent, we display this in \figref{fig:GowdyTest}\textbf{(a)} for $N=32$.\footnote{ $N = 16$ or $N=64$ are not shown because they behave the same.} 
    Additionally, the error $E_X(r)$ depends only on the radial discretization. However, the number of nodes becomes relevant at the evaluation of constraint equations, see \figref{fig:GowdyTest} \tbf{(b)}.
    Good results cannot be achieved if $N<32$. These observations agree with those from \figref{fig:HamiltonianError_Checker} and \figref{fig:MomentumError_Checker}.
    It is also important to mention that the hyperbolicity condition, Eq.\eqref{ec:hyperboliccondition}, is satisfied for $t=0.1$ used in in this experiment.
    \begin{figure}[h!]
      \begin{subfigure}{0.52\textwidth}
        \includegraphics[width=1.05\textwidth ,right]{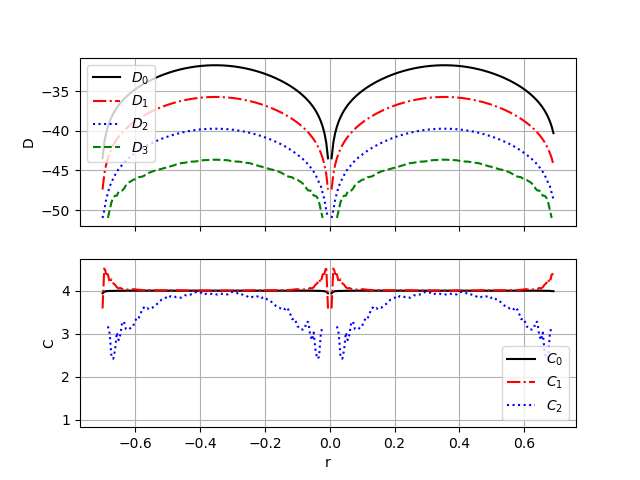}
      \end{subfigure}
    \hfill
      \begin{subfigure}{0.52\textwidth}
        \includegraphics[width=1.05\textwidth,right]{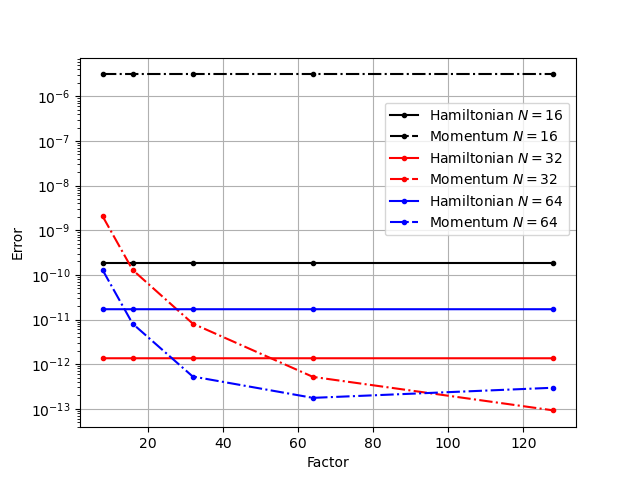}
      \end{subfigure}
      \caption{Convergence test for Gowdy metric. (\tit{left}) Convergence test for $N = 32$. (\tit{right}) Constraint violation of the solution reconstructed from the evolution process as a function of the number of nodes and number of evolution radial steps $N_{r, evol} = Factor\, N$.}
      \label{fig:GowdyTest}
    \end{figure} 
    
    \item\tbf{Test for the PFLRW metric:} Here we reproduce the analytic solutions for the perturbed PFLRW metric. This is one of the main goals of this work due to its relevance in cosmology. 
    As we can see in \figref{fig:PFLRWTest}\tbf{(a)}, the RK4 method is not convergent. 
    Among the causes of this behavior is the aliasing error. 
    We dealt with it by limiting the band-width of the functions, this is, by applying a filter that set to zero all Fourier modes beyond certain $k_{max}$, explore this filtering strategy in Appendix~\ref{appendix:aliasing_and_filter}. 
    In particular, we show results for the $(1/2)-$filter.
    % Although controlling the aliasing error do not affect the convergence test; it is crucial to obtain adequate error and constraint violation values.
    As can be seen in \figref{fig:PFLRWTest}\tbf{(b)}, the constraint violation does not converge and increases with mesh refinement.
    \begin{figure}[h!]
        \begin{subfigure}{0.52\textwidth}
          \includegraphics[width=1.05\textwidth,right]{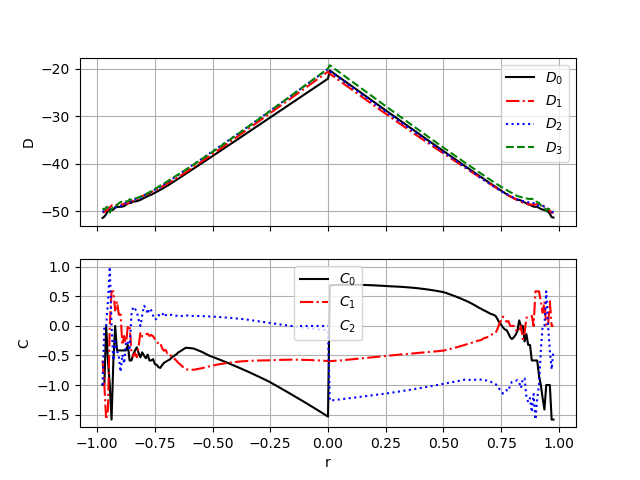}
          \caption{~}
        \end{subfigure}
        \begin{subfigure}{0.52\textwidth}
          \includegraphics[width=1.05\textwidth ,right]{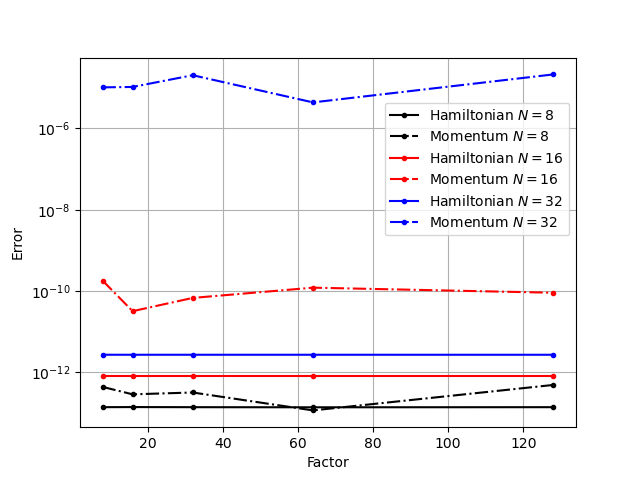}
          \caption{~}
        \end{subfigure}
          \caption{Convergence test for PFLRW metric of eq. \eqref{SPFLRW_Metric} (\tit{left}) Convergence test for $N = 32$. (\tit{right}) Constraint violation of the solution reconstructed from the evolution process as a function of the number of nodes and number of evolution radial steps $N_{r, evol} = Factor\, N$.}
          \label{fig:PFLRWTest}
        \end{figure}
        This result is further explained in Section~\ref{sec:PFLRWstability} under the light of numerical stability analysis and related to the fulfillment of the hyperbolicity condition.
\end{enumerate}

It is worth noticing that the \tit{triangle-like} shapes of convergence test plot in \figref{fig:PFLRWTest}\tbf{(a)} are due to the \tit{forward-backward} scheme that we have implemented. During the \tit{forward} (or \tit{backward}) evolution the error accumulates and grows further, this makes the solutions achieved with only one integration direction to exhibit considerably larger errors.\\

In light of \figref{fig:PFLRWTest}, we can conclude that this approach does not allow us to obtain solutions of the constraint equations for PFLRW cases. 
In the next section, we address the underlying reasons why this method does not work and which are the options to obtain stable solutions without modifying the system.

\section{Stability analysis}\label{sec:Stability}

% \CC{wrote two new paragraphs, one at the beginning and other at the end of the section to provide more context regarding the role of the fourier differentiation matrices in the error's ill behavior}

The previous section illustrated that the proposed evolution scheme leads to rapid error accumulation during radial integration for PFLRW. In this section,  by analyzing the system in the linear regime, we demonstrate that the unstable nature of the scheme is directly related to the spectrum of the discretization matrices. This, in turn, implies that the underlying problem lies in the use of the Fourier differentiation method for the approximation angular derivatives. Additionally, we show that this appears to be tied to the fulfillment of the hyperbolicity condition defined in eq. (\ref{ec:hyperboliccondition}). 

% \CC{In this section we provide a stability analysis of the linearised system and show that the error's ill behavior is a consequence of the failure of the eigenvalue distribution of the discretisation matrices to lie in the stability regions of the ODE integration methods. Additionally, we demonstrate that the spectrum of the discretisation matrices is directly related to the spectral properties of the Fourier differentiation matrices due to their block structure, suggesting that the underlying issue with the implementation is the proposed scheme for approximating derivatives in the periodic domain. Similarly, we also find that for the Gowdy metric, the numerical stability of the proposed method of lines is tied to the fulfillment of the hyperbolicity condition, i.e to the well-posedness of the given PDE system.}\\

The PDE system (\ref{drXeq}-\ref{Zeq}) of the algebraic hyperbolic formulation of the constraints is nonlinear and generally has variable coefficients that depend on the coordinates $(r,x_{1},x_{2})$.  As explained in  \cite{higham1993stiffness}, the standard approach to studying the stability of a nonlinear differential equation near an exact solution involves linearizing the system and freezing the coefficients to obtain a linear differential equation with constant coefficients. \\

Suppose that $(X^{(A)}, Y_{i}^{(A)})$ is an analytical solution of the 2+1 equations. We aim to find solutions in its vicinity by considering perturbations $(X^{(A)} + \delta X, Y_{i}^{(A)} + \delta{Y_{i}})$. To do so, we fix the freely specifiable variables derived from the 4-dimensional metric associated with the analytical solution and substitute the new solution in the equations (\ref{drXeq}-\ref{Zeq}). Disregarding nonlinear terms on the variables $( \delta X,\delta{Y_{i}})$ we obtain the following system:
\begin{align}
    \partial_{r} \delta X&= \mathcal{L}_{\widehat{\beta}}\delta X + \widehat{\alpha}\left( D_{j} \delta Y^{j} -2\delta Y^{j} \,\widehat{a}_{j} + (\delta Z-\frac{1}{2}\delta X)H^{j}_{\,j} \right) \label{eqdeltaX},\\
    \partial_{r}\delta Y_{i}&= \mathcal{L}_{\widehat{\beta}}\delta Y_{i} + \widehat{\alpha}\left( \frac{1}{2} D_{i} \delta X + D_{i}\delta Z - \delta Y_{i} H^{j}_{\,j} -\delta Z\,\widehat{a}_{i} + \frac{1}{2} \delta X\,\widehat{a}_{i} \right), \label{eqdeltaY}\\
    \delta Z&= -\left(\frac{Z^{(A)}}{X^{(A)}} + \frac{1}{2}\right)\delta X + \frac{1}{2X^{(A)}}\left(4\delta Y_{i}Y^{i \ (A)}  \right).\label{eqdeltaZ}
\end{align}
Where $Z^{(A)}$ is the original algebraic relation evaluated at the analytical solution $Z(X^{(A)},Y_{i}^{(A)})$. \\

For simplicity, we assume that the perturbations depend only on a single angular coordinate, $x_{1}$, which we denote as $x$. By defining the 1-dimensional angular grid $\{i\Delta x\}_{i=1}^{N}$ in the interval $[0,2L]$, with spacing $\Delta x = 2L/N$ and approximating all angular derivatives using the Fourier collocation method described in Section \ref{section:numerical_approach}, the previous system can be written as a coupled system of ordinary differential equations,

\begin{equation}
    \partial_{r}\boldsymbol{U}= \boldsymbol{L}U,
    \label{semidiscretizedeq}
\end{equation}
where,
\begin{equation}
    \boldsymbol{U}^{T}= (\delta X(r,x_{1}) ,\ldots, \delta X(r,x_{N}), \delta Y_{1}(r,x_{1}), \ldots, \delta Y_{1}(r,x_{N}), \delta Y_{2}(r,x_{1}), \ldots \delta Y_{2}(r,x_{N})),
\end{equation}
and $\boldsymbol{L}$ is the $3N \times 3N$ semi-discretization matrix, that couples the grid values. The matrix entries will be fixed because of the freezing coefficients process.\\

The previous system can be integrated in the radial direction using either explicit schemes like RK4 or implicit schemes like implicit Euler or Crank-Nicholson methods. Following the radial discretization and solving for $\bsym{U}^{n+1}$ if necessary, equation \ref{semidiscretizedeq}, becomes

\begin{equation}
    \bsym{U}^{n+1}= \boldsymbol{A_{\Delta r}}\bsym{U}^{n},
    \label{discretizedapproxeq}
\end{equation}
where $\boldsymbol{U}^{n}=\boldsymbol{U}(r_{n})$ and we call $\boldsymbol{A_{\Delta r}}$ the \tit{full discretization matrix}.

\subsection{Stability and eigenvalue criteria} 

A discretized approximation to a PDE, as the one presented in equation \ref{semidiscretizedeq}, is stable if the powers of the full discretization matrix are uniformly bounded as
\begin{align}
\label{eq:LaxStabilityCondition}
    \| (\bsym{A}_{\Delta_{r}})^{n}\|\leq C(n\Delta r),
\end{align}
for a fixed function $C(r)$, for all $n$ and $\Delta r, \Delta x \rightarrow 0$. 
Here $\|\cdot\|$ denotes any operator norm induced by a vector norm. 
For the remainder of this section, it will specifically refer to the matrix norm induced by the $L^{2}$ norm.
The fulfillment of this condition is known as Lax-stability, and because of the Lax Equivalence Theorem, it implies the convergence of the method (see Chapter 7 of \cite{trefethen2005spectra}). \\ 

Although strong and general, this criteria is difficult to verify.
Since it requires the evaluations of the discretization matrices $\bsym{A}_{\Delta_{r}}$, it may become impossible to evaluate for implicit methods.
In practice, it is possible to determine the (eigenvalue-)stability of a scheme based on the method of lines by analyzing the spectrum of the spatial discretization matrix $\bsym{L}$ for a fixed mesh size.
\begin{definition}[Eigenvalue stability]
\label{def:EigenvalueStability}
    The method of lines of eqs.\eqref{semidiscretizedeq} and \eqref{discretizedapproxeq} is eigenvalue-stable if the eigenvalues of the discretized spatial operator $\bsym{L}$, scaled by $\Delta r$, fall within the stability region of the \tit{time-discretization} method\,\footnote{See \cite{Trefethen} for the definition of stability regions in ODE methods}.
\end{definition}
Although, in general, the eigenvalue-stability is not sufficient to guarantee the convergence of the method as $\Delta t \to 0$ (it is weaker than Lax-stability), it is a necessary condition. 
Additionally, for certain schemes, as those where $\bsym{L}$ is a normal matrix\,\footnote{A matrix $A$ is normal if it commutes with its conjugated transpose $A^*$, this is, $AA^*=A^*A$.}, eigenvalue-stability implies Lax-stability (see \cite{trefethen2005spectra,mortonlax}). 
This makes of this criteria a powerful tool to practically evaluate the stability of a method of lines.\\

% \AC{State as theorem (or props+theo): the method of lines discretization of the linear system with frozen coefficients based on Fourier differentiation method is always unstable for the FLRW (PFLRW?) spacetime}\\
% \AC{prop1: (remind from some reference) there is no consistent Runge-Kutta ODE method whose stability region includes any interval real interval of the form $(0,a)$ for small positive $a$.\\
% See for example Sec 35 of \cite{Butcher2016}. The consistency condition of RK methods reads $\sum_i b_i = 1$. And the stability function is $R(z) = 1+z\mbf{b}^T(I - zA)^{-1}\mbf{e}$. This means that for small $z$, $R(z) = 1+z + O(z^2)$ (use Taylor expansion $(I - z A)^{-1} = I + zA + z^2 A^2 +\cdots$ and $\mbf{b}^T\mbf{e}=1$).\\
% Therefore for small real positive $z$, $|R(z)|>1$ \ie $(0,a)$ is not in the stability region for small real positive number $a$.}
To make use of it, we need to determine the stability region of the ODE methods we use for the \tit{time-discretization}. 
In this work, every ODE method used falls within the category of Runge-Kutta (RK) methods.  
This means that (see \cite{Butcher2016} and \cite{Stoer_Bulirsch}), given an ODE of the form $y(t) = f(t,y)$, it can be written as 
\begin{equation*}
    y^{n+1} = y^n + h \sum_{i=1}^s b_i k_i ~ \text{ with } ~
    k_i = f\left(t_n + c_i\, h, \,y^n + \sum_{j=1}^sa_{ij}k_j\right),
\end{equation*}
and $\bsym{A} = \{a_{ij}\}_{i,j = 1}^s$, $\bsym{b} = \{b_i\}_{i=1}^s$ and $\bsym{c} = \{c_i\}_{i=1}^s$ the coefficients of the method, where the integer $s$ denotes the number of \tit{stages} of the method.
The method is consistent if the condition 
\begin{equation*}
    \sum_{i=1}^s b_i = \mbf{b}^T\mbf{e}= 1
\end{equation*}
holds, where $\mbf{e} = (1,1, \cdots, 1)$ is a \tit{vector} of $s$ components. 
Additionally, their stability region is the set 
\begin{equation*}
    S = \{z\in\mathbb{C}:|R(z)|\leq1\},
\end{equation*}
where
\begin{equation}
    R(z) = 1+z\mbf{b}^T(I - z\bsym{A})^{-1}\mbf{e}.
\end{equation}
For these methods, following Sec 35 of \cite{Butcher2016}, we can recall the following result
\begin{prop}
\label{prop:RKstability}
    There is no consistent Runge-Kutta ODE method whose stability region includes any real interval of the form $(0,a)$ for small positive $a$.
\end{prop}
\begin{proof}
Using the Taylor the expansion $(I - z \bsym{A})^{-1} = I + z\bsym{A} + z^2 \bsym{A}^2 +\cdots$ and the consistency condition, we see that $R(z) = 1+z + O(z^2)$.
Therefore, for small real positive $z=a$, $|R(a)|>1$ \ie\, $(0,a)$ is not in the stability region for small real positive number $a$.
\end{proof}
Note that this additionally implies that any complex number given by $z=a+i b$ with small enough $a$ lies outside of the stability region. 
From here, we can immediately see the following consequence:
% \AC{prop2: $L$ eigenvalues take values on the positive real axis.}\\
\begin{pcoro}
\label{coro:SemidiscretizationRealStability}
    If the eigenvalues of the semi-discretization matrix $\bsym{L}$ from the semi-discretized system of eq.(\ref{semidiscretizedeq}) have small positive real part, then the numerical scheme is unstable.
\end{pcoro}

However, as mentioned previously, eigenvalue-stability is necessary but not sufficient to ensure Lax-stability (and therefore convergence) for cases where the spatial discretization matrix of the scheme is \tit{non-normal} (see \cite{trefethen2005spectra, mortonlax,reddy1992stability}).
% \AC{Enunciate general theorem from Chapter 7 of \cite{trefethen2005spectra} then state criteria},\\
For such cases a new kind of stability for the method of lines was introduced by Reddy and Trefethen in \cite{reddy1992stability} consists on studying the $\epsilon-$pseudospectrum of the semi-discretization matrix $\bsym{L}$.
The $\epsilon-$pseudospectrum intends to characterize how the spectrum of a matrix behaves under perturbations and is defined as 
\begin{definition}[$\epsilon-$pseudospectrum]
    The $\epsilon-$pseudospectrum of a matrix $L$ is defined as the region
\begin{equation}
    \Lambda_{\epsilon}(\bsym{L})= \{ z \in \mathbb{C} : \| (zI - \bsym{L})^{-1} \|_{2} >\epsilon \}.
\end{equation}
\end{definition}
In \cite{trefethen2005spectra} (Thrm. 32.2) it is stated that if the distance between the $\epsilon-$pseudospectrum of the semi-discretization matrix lies within a distance $O(\epsilon)+O(\Delta r)$ of the stability region as $\epsilon \rightarrow 0$ for sufficiently small $\Delta r$, then the method of lines is stable in this new sense.
This motivates the following definition: 
\begin{definition}[$\epsilon$-stability of a Method of Lines]
\label{def:EpsilonStability}
    The method of lines of a \tit{time}-dependent system with semi-discretization matrix $\bsym{L}$ and \tit{temporal} discretization stability region $S$ is $\epsilon$-stable if 
    \begin{equation}
        \dist(\Lambda_{\epsilon}(\bsym{L}),S) := \max_{(\mu_{\epsilon},x)\in\Lambda_{\epsilon}(\bsym{L})\times S} \dist(\mu_{\epsilon},x) = \mathcal{O}(\epsilon) + \mathcal{O}(\Delta t) 
    \end{equation}
    as $\epsilon,\Delta t \to 0$, where $\Delta t$ is the step-size of the \tit{time}-discretization.
\end{definition}
This kind of stability is again weaker than Lax-stability. 
In general, it implies a bound on the full-discretization matrix similar to eq.\eqref{eq:LaxStabilityCondition} where the function $C(\cdot)$ is not guarantee to converge to zero if $\Delta t \to 0$ (or equivalently $n\to\infty$) but behave as $\min(N,n)$ for $0\leq n\Delta t \leq T$ for some $T \in (0,\infty)$ and $N$ the number of nodes in the \tit{spatial}-discretization.
However, as discussed in \cite{trefethen2005spectra} Section 32, 
this is usually enough to obtain convergence for a large subclass of problems. 
In particular, this holds for those problems with smooth initial data, which are the class of interest in the present work.

\subsection{FLRW and PFLRW spacetimes stability}
\label{sec:PFLRWstability}

To elucidate the stability issues of the MOL for the PFLRW metric, we initially focus on the unperturbed case. Since FLRW AHF system has constant coefficients, we can analytically find the eigenvalues of the spatial discretization matrix. We assume for the analysis that the perturbations around the analytical solution do not depend on the angular variable $x_{2}$. \\

\paragraph{Unperturbed case.} Writing down the explicit form of the AHF system for the FLRW metric, we obtain,
\begin{alignat}{2}
\label{RWahfEquations}
    \partial_{r} \delta X &= \frac{1}{a(\eta)} \partial_{x} \delta Y_{1}, \quad &&x\in [0,2L], \quad r \in [0,2L];\\
    \partial_{r} \delta Y_{1} &= -\frac{a(\eta)}{2} \partial_{x}\delta X, \quad 
    &&x\in [0,2L], \quad r \in [0,2L]; \\
    \partial_{r} \delta Y_{2} &= 0, \quad &&x\in [0,2L], \quad r \in [0,2L] . 
    \label{FLRW2+1}
\end{alignat}

After the semi-discretization process, the system can be written in the form of the equation \ref{semidiscretizedeq}. Where the semi-discretization matrix is given by
\begin{align}
 \mathbf{L} &= \left(\begin{array}{ c | c | c }
   0 & \frac{1}{a(\eta)} D& 0\\
    \hline
    -\frac{a(\eta)}{2} D  & 0 & 0 \\
    \hline
    0 & 0 & 0
  \end{array}\right). \label{eq:FLRWsemidismatrix}
\end{align}
Here $D$ denotes the Fourier first-order differentiation matrix of size $N \times N$. The corresponding non-trivial eigenvalues of the previous matrix are given by
\begin{align}
    \label{eq:UnperturbedEigenvalues}
    \lambda= \pm \frac{\pi}{\sqrt{2}L}k \qquad k = -\frac{N}{2}+1, \ldots, \frac{N}{2}-1.
\end{align}

 \begin{figure}[h!]
    \centering
    \subfloat[\centering]
    {{\includegraphics[width=6.5cm]{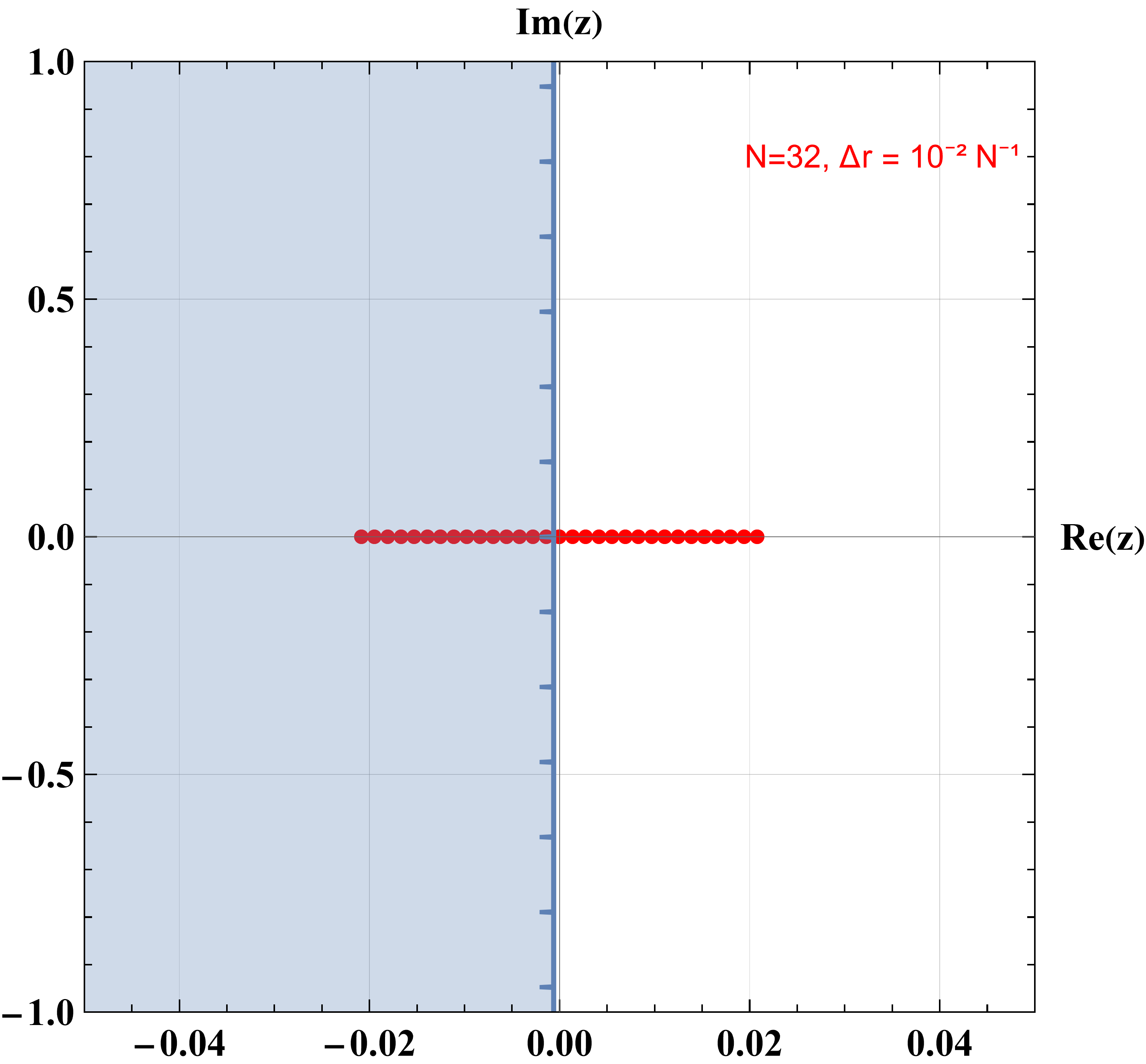} }}%
    \subfloat[\centering]
    {{\includegraphics[width=6.5cm]{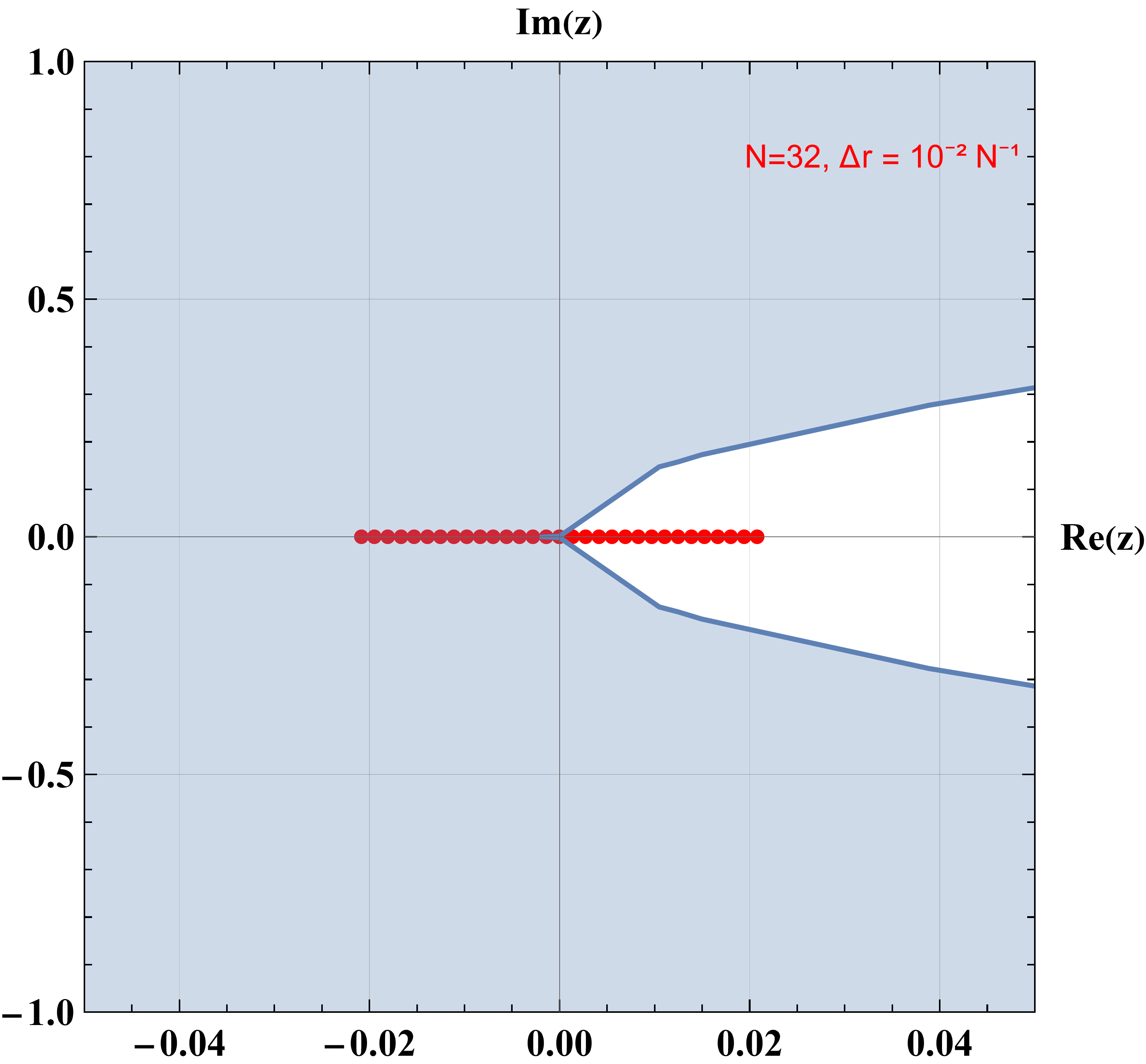} }}%
    \caption{Eigenvalues of $\Delta r$ times the Matrix $\bsym{L}$,  for $L=0.5$, superimposed on the stability region of (from left to right) Crank Nicholson and Implicit Euler. The eigenvalues were scaled by a factor $\theta=10^{2}$ for visualization purposes.}
    \label{fig:StabilityFLRW}
\end{figure}

As illustrated in figure \ref{fig:StabilityFLRW}, half of the scaled eigenvalues of $\bsym{L}$ lie outside the stability region of all the integration schemes considered in this work. Consequently, since the eigenvalue criterion is a necessary condition for stability, regardless of the radial step $\Delta r$, we conclude that the method of lines is not stable for this system.\\

\paragraph{Perturbed case.} In the context of the PFLRW system, the stability analysis is more complicated than in FLRW due to the metric tensor depending on both angular and radial variables. This results in a more complicated structure for the spatial discretization operator that is not tractable analytically.\\

Nonetheless, since we are working with a potential with amplitude of order $10^{-8}$, the spectrum of the semi-discretization matrix can be expected to resemble the FLRW case. To verify this claim, we perform a numerical exploration of the spectrum of the discretization matrix rather than an analytical computation like the one presented for the non-perturbed metric. For the following analysis, we assume a potential of the form $\phi= \phi_{0}(( \sin(\pi r/L)  + \sin(\pi x/L))$.\\

To compute the eigenvalues of the discretization matrix, we rewrite the linearized AHF system (\ref{eqdeltaX}-\ref{eqdeltaZ}) in a schematic
form that allows us to identify the blocks of $\bsym{L}$ as with the previous system,
\begin{align}
    \partial_{r} U_{i}= h_{i}(x,r) \partial_{x}U_{i} + b_{~i}^{j}(x,r) \partial_{x} U_{j} +
c_{i}(x,r) U_{i} + m_{~i}^{j}(x,r)U_{j} \qquad i,j=1,2,3. \label{eq:GeneralForm}
\end{align}
Here, $U = (X,Y_{1},Y_{2})$ and the domain of definition of each component function is the region $[0,1]\times [0,1]$.\\

To perform the same spectrum analysis as before, we need the system to be independent of the coordinate variables $r$ and $x_{1}$. Since the PDE for the PFLRW metric does not fulfill this condition, we employ the strategy of frozen coefficients. 
This strategy involves fixing the values of the system coefficients on the 2-dimensional mesh $\{j\Delta r, i\Delta x\}_{i,j=1}^{N}$. In this work, we consider two different methods for freezing the coefficients: the first method uses their average value over the mesh, and the second uses their maximum absolute value within the mesh.\\

The semi-discretization matrix for the system with frozen coefficients is given by
\begin{align}
 \bsym{L} &= \left(\begin{array}{ c | c | c }
   h_{1} D +c_{1} I & b_{1}^{1}D + m_{1}^{1} I& b_{1}^{2}D + m_{1}^{2}I\\
    \hline
    h_{2} D +c_{2} I & b_{2}^{1}D + m_{2}^{1}I & b_{2}^{2}D + m_{2}^{2}I\\
    \hline
    h_{3} D +c_{3} I & b_{3}^{1}D + m_{3}^{1}I & b_{3}^{2}D + m_{3}^{2}I
  \end{array}\right). \label{eq:semidiscretematrix}
\end{align}
Since, the perturbation $\phi$ controlled by the parameter $\phi_0$, this matrix reduces to eq.(\ref{eq:FLRWsemidismatrix}) when $\phi_0\to 0$. 
This allows us to write it as 
\begin{equation*}
    \bsym{L} = \overline{\bsym{L}} + \phi_0\delta\bsym{L} + \mcal{O}(\phi_0^2), 
\end{equation*}
where $\overline{\bsym{L}}$ is the semi-discretization matrix of the unperturbed FLRW system eq.(\ref{eq:FLRWsemidismatrix}).
Therefore, the eigenvalues of this matrix, up to first order in $\phi_0$, can be written as 
\begin{equation*}
    \lambda(\phi_0) = \lambda_0 + \phi_0\lambda_1,
\end{equation*}
where $\lambda_0$ take values on the list of eq.(\ref{eq:UnperturbedEigenvalues}) and $\lambda_1$ is given by $\lambda_1 = w_0^T\delta\bsym{L}v_0$ with $w_0$ and $v_0$ the left and right eigenvectors of $\overline{\bsym{L}}$ respectively if the eigenvector normalization $ w_0^Tv_0=1$ is chosen.
In consequence, the eigenvalues of the semi-discretization matrix of the PFLRW metric are small deviations of the eigenvalues of the unperturbed case. 
Thus, they take values outside of the convergence region of any RK method which makes the scheme unstable.\\

With this computation of the eigenvectors and the results of Prop.\ref{prop:RKstability} and Coro.\ref{coro:SemidiscretizationRealStability}, we have proved the following result:
\begin{theorem}\label{teorema}
    The method of lines based of Fourier differentiation described in Section(\ref{section:numerical_approach}) applied to the Algebraic Hyperbolic Formulation of the Constraint Equations of Section(\ref{section:AHF_presentation}) is unstable for scalar perturbations of FLRW spacetimes of the form eq.(\ref{SPFLRW_Metric}).
\end{theorem}
Although this is a pessimistic result, removing any of the hypothesis might improve the result. 
Therefore, it gives us a path to obtain solutions. 
In particular, changing the metric, as is done in Section~\ref{sec:GowdyStability} or modifying the PDE system, as we discuss in Section~\ref{section:new_id}, might allow stable solutions. \\ 

\begin{figure}[ht]
    \centering
    \subfloat[\centering]
    {{\includegraphics[width=6.5cm]{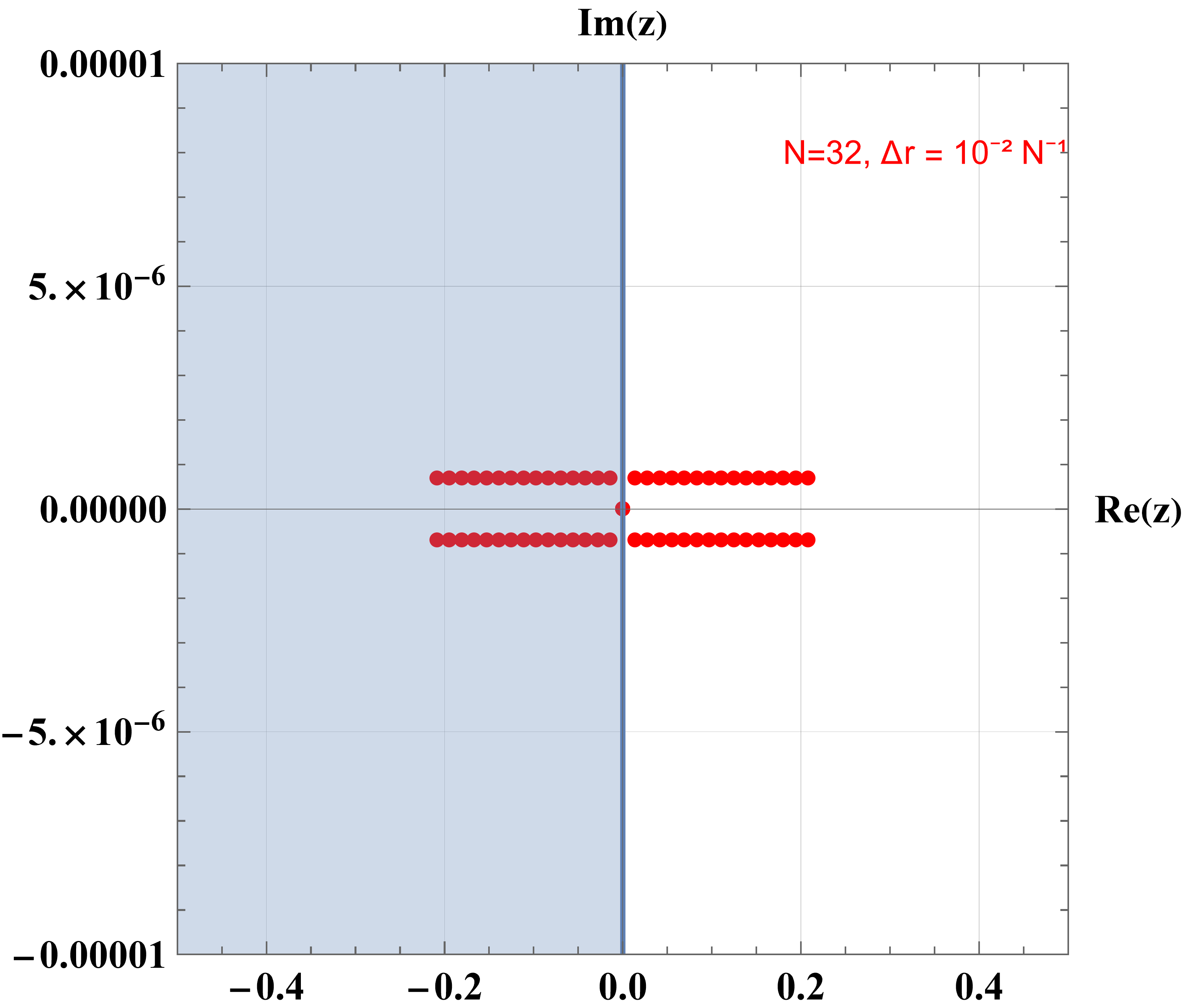} }}%
    \subfloat[\centering]
    {{\includegraphics[width=6.5cm]{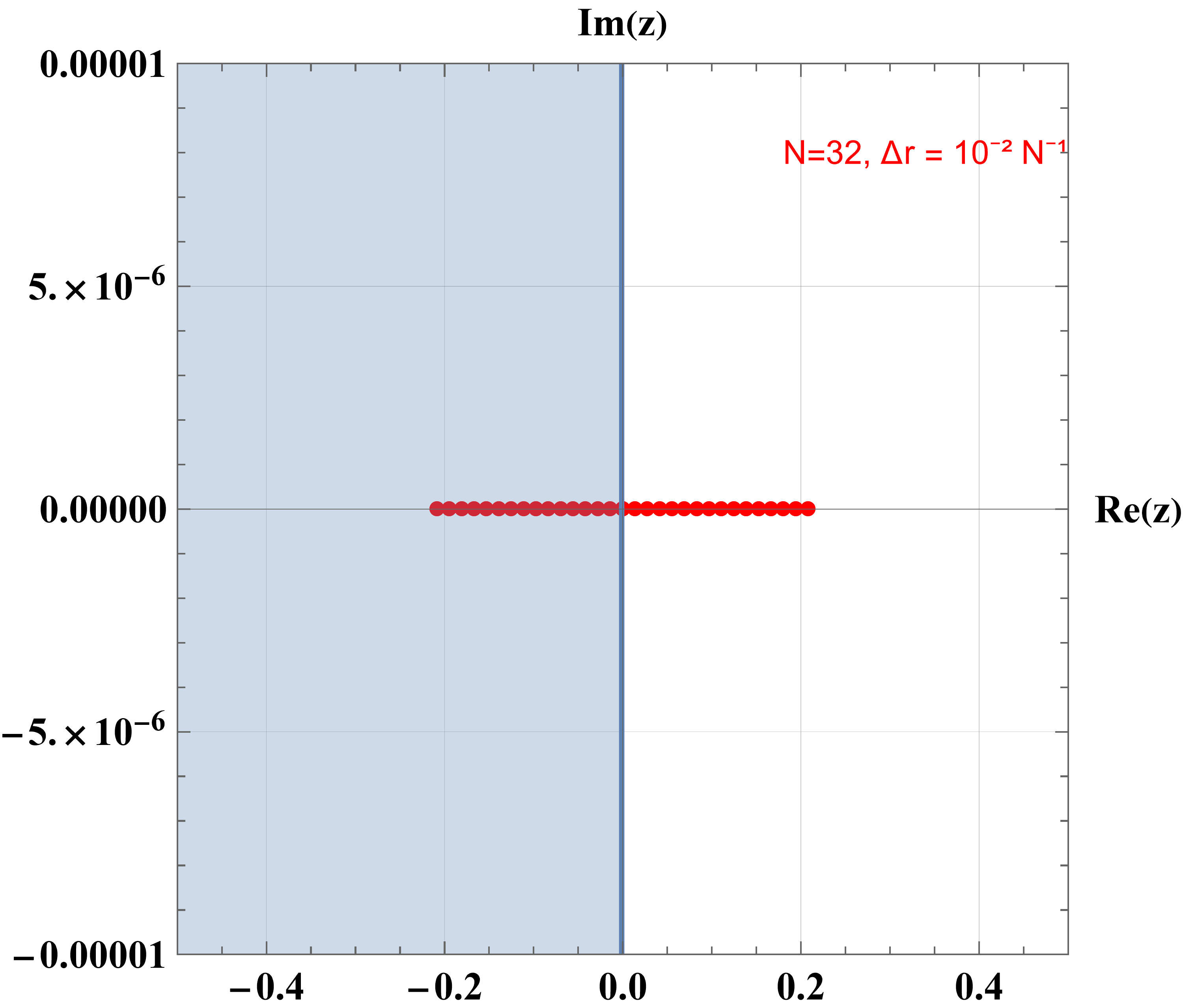} }}%
    \caption{Left: Eigenvalues of $\Delta r$ times the matrix $\bsym{L}$ with coefficients frozen by taking their maximum in the mesh. Right: Eigenvalues of $\Delta r$ times the matrix $\bsym{L}$ with coefficients frozen by taking their average in the mesh. In both cases, the spectrum is superimposed on the stability region of CN and scaled by a factor $\theta=10^{2}$ for visualization purposes.}
    \label{fig:StabilityPFLRW}
\end{figure}

Figure \ref{fig:StabilityPFLRW} shows the distribution of the eigenvalues of the angular discretization operator when the coefficients in equation \ref{eq:GeneralForm} are frozen using the two different methods mentioned above, for $\Delta r= 10^{-2} N^{-1}$. In both cases, the numerically calculated eigenvalues fall outside the stability region, regardless of the chosen radial step, just as in the unperturbed case. This result indicates that, similar to the FLRW system, the behavior of the spectrum of the angular discretization matrix is dominated by the spectrum of the Fourier differentiation matrix.\\

\begin{figure}[ht]
    \centering
    \subfloat[\centering]
    {{\includegraphics[width=6.5cm]{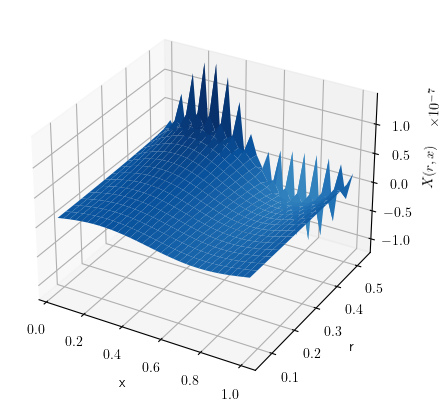} }}%
    \subfloat[\centering]
    {{\includegraphics[width=6.5cm]{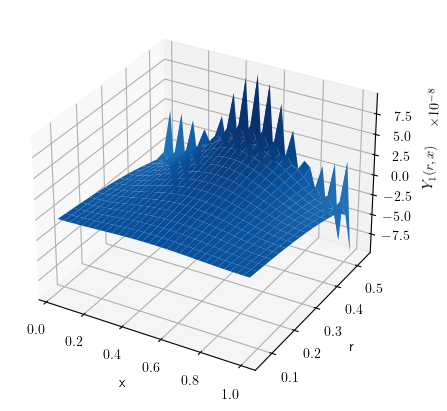} }}%
    \caption{behavior of the numerical solution of the linearized AHF equations using CN for PFLRW spacetime under initial condition modifications of the first component, for N=32 nodes and $\Delta r= 10^{-2}/N$. \textbf{(a)} Solution for the first component $\delta X(r,x)$. \textbf{(b)} Solution for the second component $\delta Y_{1}(r,x)$.}
    \label{fig:WaterfallsPFLRW}%
\end{figure}

To visualize the previous results, in figure \ref{fig:WaterfallsPFLRW}, we present  both components of the numerical solution of the linearized AHF equations for an initial condition for $(\delta X, \delta Y_{1}, \delta Y_{2})$ of the form
\begin{equation}
\begin{split}
    \delta X(0,x) &= a +a \sin( 2\pi x), \quad a=10^{-8}. \\
    \delta Y_{1}(0,x)&= \delta Y_{2}(0,x)=0
    \label{eq:InitialCondition}
\end{split}
\end{equation}
    
 As expected from the stability analysis of the linearized equations, the system is numerically unstable, this instability is reflected by the rapidly growing oscillations that appear as the radial variable increases in  figure \ref{fig:WaterfallsPFLRW}.

\subsection{Gowdy spacetime stability}
\label{sec:GowdyStability}

Here, we illustrate how this scheme can still be stable if the spacetime class is changed but also how the stability may strongly depend on the parameters of the metric.
For the Gowdy spacetime eq.\eqref{Gowdy_Metric}, the explicit form of the linearized AHF PDE system is given by:
\begin{alignat}{2}
\partial_{r} \delta X &= \left(\frac{\hat{\alpha}e^{P}}{t}\right) \ \partial_{x} \delta Y_{1}, \quad &&x\in [0,1], \quad r \in [0,1], \label{eq:deltaXGowdy}\\
    \partial_{r} \delta Y_{1} &= - \left(\frac{\hat{\alpha} Z^{(A)}}{X^{(A)}}\right) \ \partial_{x}\delta X, \quad 
    &&x\in [0,1], \quad r \in [0,1], \label{eq:deltaY1Gowdy}\\
    \partial_{r} \delta Y_{2} &= 0, \quad &&x\in [0,1], \quad r \in [0,1] \label{eq:deltaY2Gowdy}.
\end{alignat}
Where, $Z^{(A)}$, $X^{(A)}$ are defined in the same way as in equations (\ref{eqdeltaX}-\ref{eqdeltaZ}) and $\hat{\alpha}$ is the two-dimensional lapse function given by $\hat{\alpha}=\left(\frac{e^{Q}}{t} \right)^{\frac{1}{4}}$.\\
The spatial discretization matrix for this system is given by

\begin{align}
  \bsym{L}(r) &= \left(\begin{array}{ c | c | c }
   0 & \left(\frac{\hat{\alpha}e^{P}}{t}\right) D& 0\\
    \hline
    -\left(\frac{\hat{\alpha} Z^{(A)}}{X^{(A)}}\right) D  & 0 & 0 \\
    \hline
    0 & 0 & 0
  \end{array}\right) \label{eq:Gowdymatrix}.
\end{align}

Similarly to the FLRW case, we can compute the eigenvalue analytically

\begin{align}
\lambda_{k}(r,t) = \pm \hat{\alpha} \left(\frac{e^{P}}{t}\frac{Z^{(A)}}{X^{(A)}}\right)^{\frac{1}{2}} \frac{\pi k}{L}, \qquad  k = -\frac{N}{2}+1, \ldots, \frac{N}{2}-1 . \label{eq:eigenvaluesGowdy}
\end{align}

Note that these eigenvalues and their distribution in the complex plane will depend on both the time parameter and the radial variable. Since $\frac{e^{P}}{t} > 0$, the eigenvalues are either pure imaginary or real numbers depending on the sign of the quotient $\frac{Z^{(A)}}{X^{(A)}}$. Alternatively, because the sign of the quotient is the same as the sign of the product $X^{(A)}Z^{(A)}$, we can focus on the latter. By doing this, we can relate the numerical stability of the scheme to the hyperbolicity condition in \ref{ec:hyperboliccondition}.

\begin{figure}[h!]
    \centering
    \includegraphics[scale=0.7]{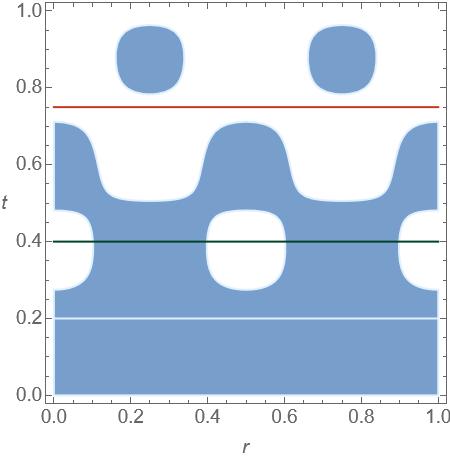}
    \caption{Graph of the inequality $XZ$ in the domain $[0,1] \times [0,1]$. 
    The blue region indicates where the hyperbolicity condition is satisfied ($XZ<0$). 
The horizontal lines illustrate the three possibilities for the parameter t: the white line at $t = 0.2$ represents a value of t for which the hyperbolicity condition is always satisfied in all the radial domain $[0, 1]$.
Whereas the green line at $t = 0.4$ represents a value for which the condition is satisfied depending on the value of the variable $r$, and the red line at $t = 0.75$ represents a value for which the condition is never satisfied.
}
    \label{fig:stabilityGowdy}
\end{figure}

Figure \ref{fig:stabilityGowdy} illustrates the behavior of the product $X^{(A)}Z^{(A)}$ in the domain $[0,1]^{2}$. In this domain, there are three possibilities for the sign of the product depending on the parameter $t$. This product can be positive or negative for all values of $r$ in the interval $[0,1]$, or its sign can change with $r$. For the cases where the sign does not change in the interval $[0,1]$, it is possible to derive some conclusions about the stability of the AHF system.

\subsubsection{Not stable case:}
\begin{figure}[h!]
    \centering
    \subfloat[\centering]
    {{\includegraphics[width=6cm]{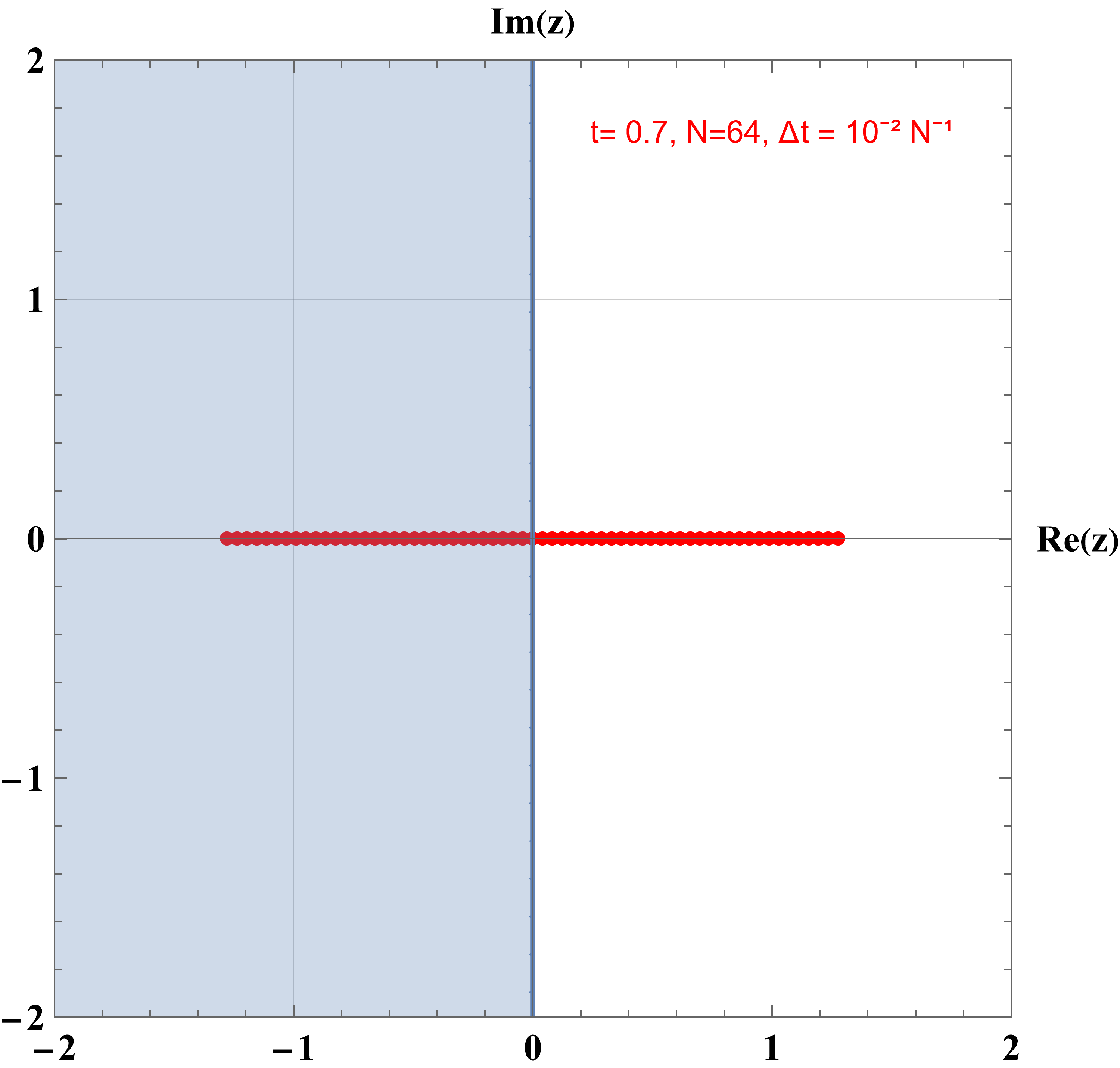} }}%
    \subfloat[\centering]
    {{\includegraphics[width=6cm]{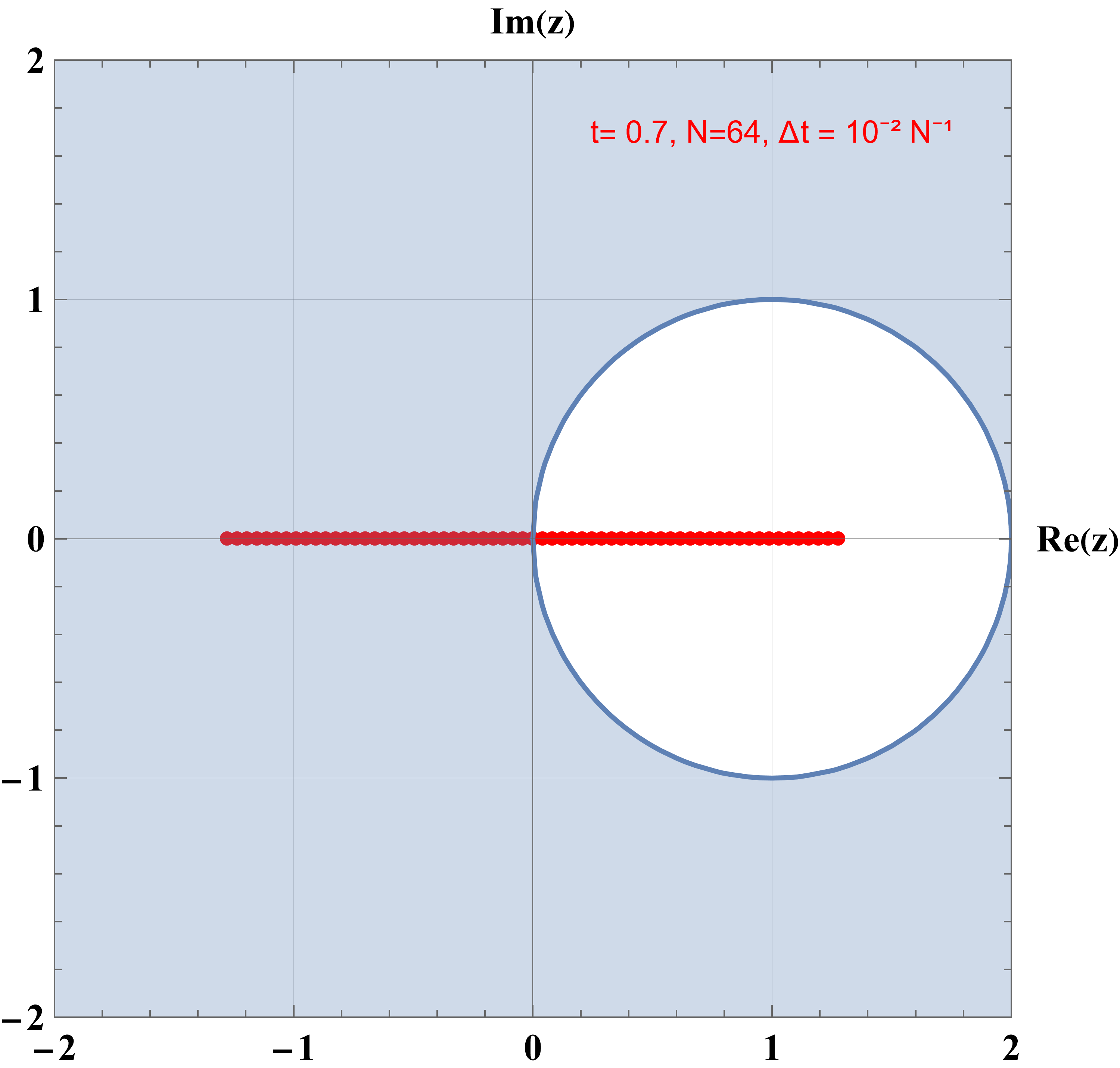} }}%
    \caption{ On the left, Eigenvalues of $\Delta r$ times the Matrix $\bsym{L}$, with coefficients frozen by taking their average in the mesh, superimposed on the stability region of Crank Nicholson. On the right, the eigenvalues are superimposed on the stability region of the Implicit Euler method. The eigenvalues were scaled by a factor $\theta=10^{2}$ for visualization purposes.}%
    \label{fig:stabilityregionsgowdyunstable}%
\end{figure}

If the product $X^{(A)}Z^{(A)}$ is always positive, the eigenvalues described by eq. \ref{eq:eigenvaluesGowdy} will be real and symmetrically distributed with respect to the imaginary axis regardless of the value of $r$. Consequently, some of the eigenvalues will lie in the right half plane, outside the stability region of any RK scheme throughout the evolution process. Figure \ref{fig:stabilityregionsgowdyunstable} shows the eigenvalues of the discretization matrix $\bsym{L}$ obtained after freezing the coefficients by taking their average over the 2-dimensional mesh across the region $[0,1]^{2}$.\\

\begin{figure}[h!]
    \centering
    \subfloat[\centering]
    {{\includegraphics[width=6.5cm]{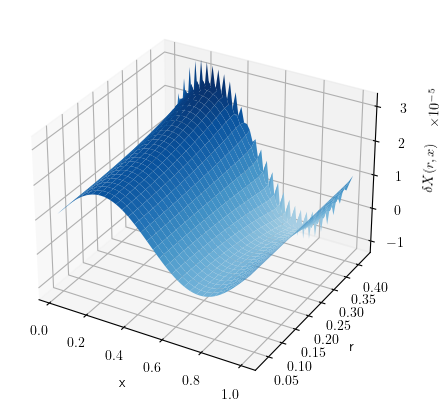} }}%
    \subfloat[\centering]
    {{\includegraphics[width=6.5cm]{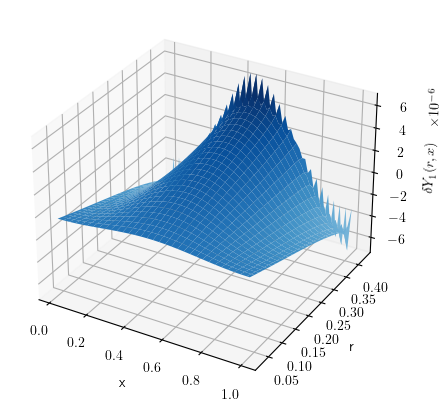} }}%
    \caption{behavior of the numerical solution of the AHF equations with CN for Gowdy spacetime under initial condition modifications of the first component, for $N=64$ nodes, $\Delta r= 10^{-2}/N$ and $t=0.7$ \textbf{(a)} Solution for the first component $\delta X(r,x)$. \textbf{(b)} Solution for the second component $\delta Y_{1}(r,x)$.}
    \label{fig:Waterfallsgowdyunstable}%
\end{figure}

Figure \ref{fig:Waterfallsgowdyunstable}, shows numerical results for $\delta X(r,x)$ and $\delta Y_{1}(r,x)$ for an initial condition $(\delta X, \delta Y_{1}, \delta Y_{2})$ of the form given in eq. \eqref{eq:InitialCondition}. The numerical solution obtained is unstable, and strong oscillations appear as the radial variable increases.
 
\subsubsection{Stable case:}
\begin{figure}[h!]
    \centering
    \subfloat[\centering]
    {{\includegraphics[width=6cm]{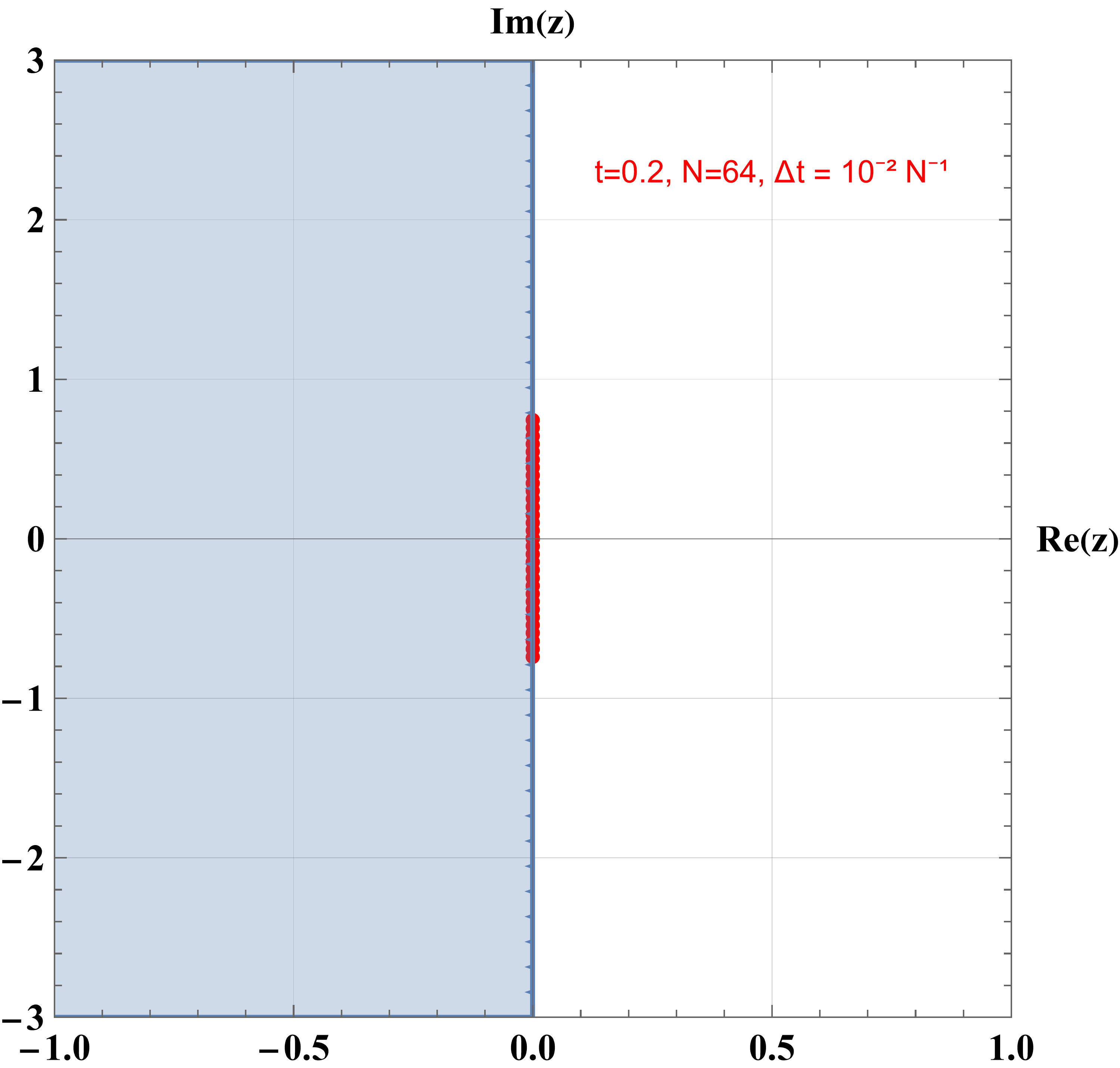} }}%
    \subfloat[\centering]
    {{\includegraphics[width=6cm]{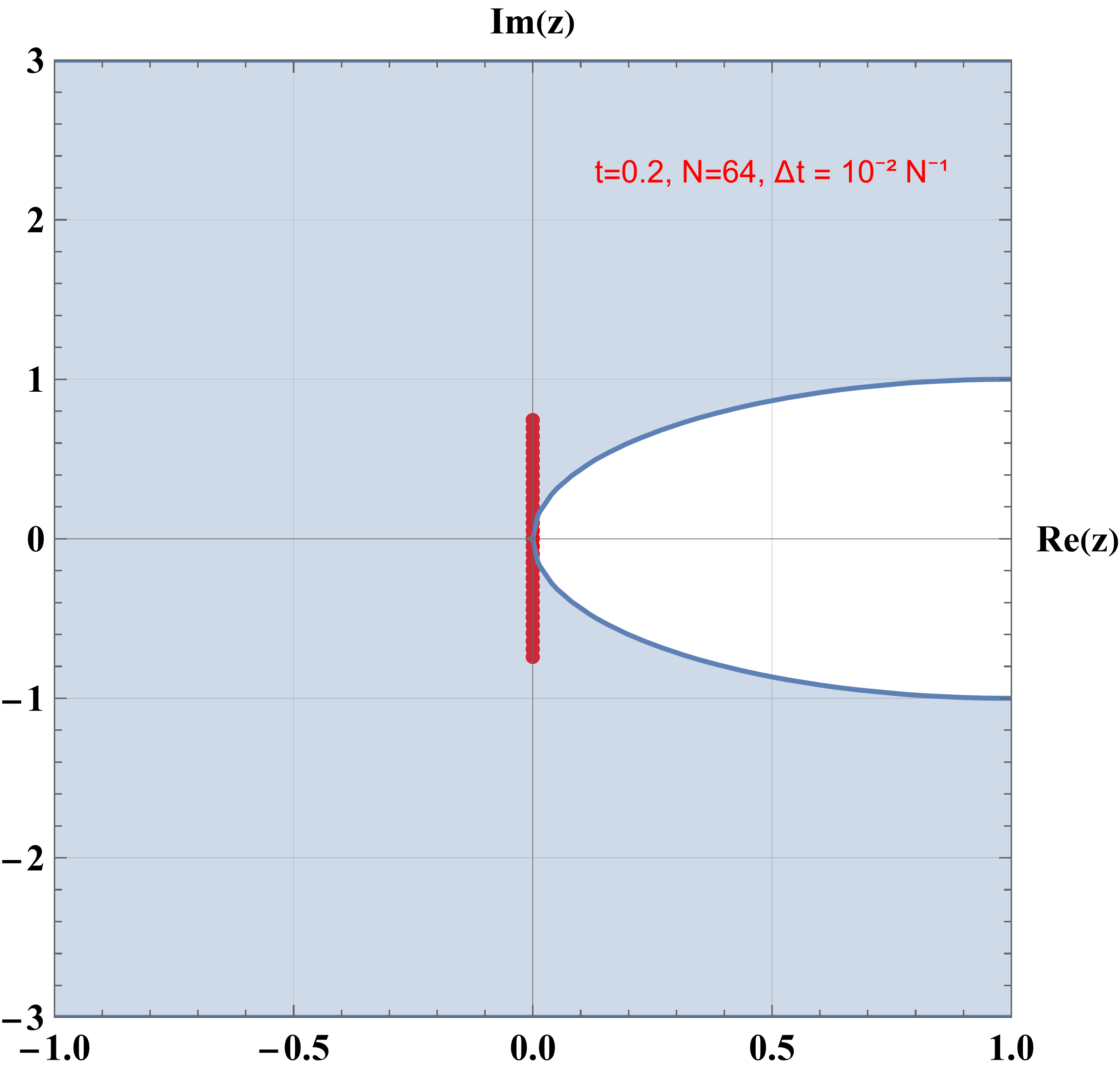} }}%
    \caption{ On the left, eigenvalues of $\Delta r$ times the matrix $\bsym{L}$,  for $L=0.5$ and $t=0.2$, superimposed on the stability region of Crank-Nicholson and on the right superimposed on the stability region of Implicit Euler. The eigenvalues were scaled by a factor $\theta=10^{2}$ for visualization purposes.}%
\label{fig:GowdyEigenvaluesStable}
\end{figure}

If the product $X^{(A)}Z^{(A)}$ is negative for all $r$, the eigenvalues in equation \ref{eq:eigenvaluesGowdy} lie on the imaginary axis and stay within the stability region of both Crank-Nicholson and Implicit Euler, independent of the number of nodes in the angular axis $N$. Figure  \ref{fig:GowdyEigenvaluesStable} shows the eigenvalues of the semi-discretization matrix with coefficients frozen by taking their average over the 2-dimensional mesh for $N=64$ nodes.\\

As indicated previously, this is not a sufficient condition to ensure the stability of the numerical scheme, since the semi-discretization matrix $\bsym{L}$ is not normal. To obtain more decisive conclusions regarding the stability of the discretization, we analyze the distance between the $\epsilon$-pseudospectrum of the frozen coefficient semi-discretization matrix \ref{eq:Gowdymatrix}, obtained by taking the average value of the coefficients over the mesh, and the stability region of the Crank-Nicholson method.

\begin{figure}[h!]
    \centering
    \subfloat[\centering]
    {{\includegraphics[width=7.2cm]{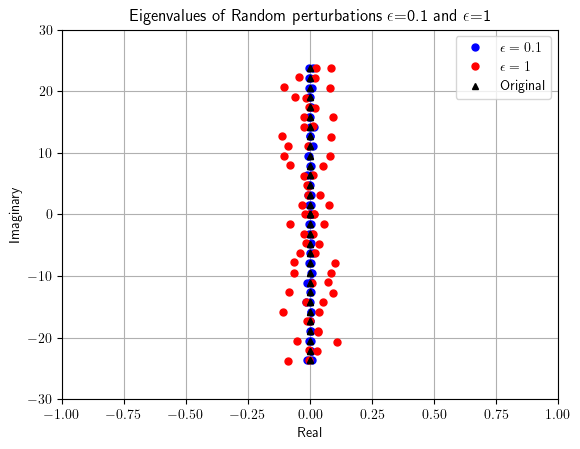} }}%
    \qquad
    \subfloat[\centering]
    {{\includegraphics[width=7.2cm]{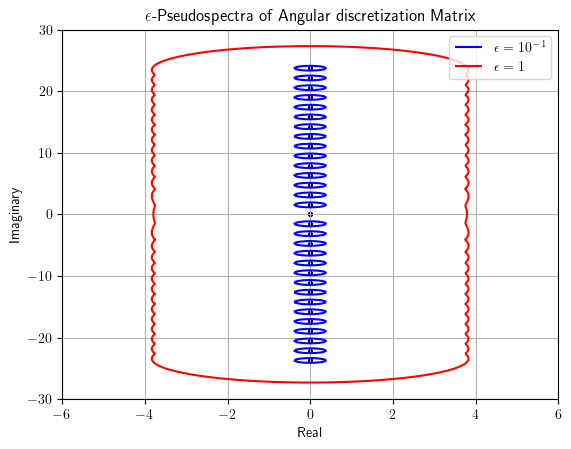} }}%
    \caption{On the right, eigenvalues of random perturbations $\mathbf{L}+E$, $\|E\|_{2}= \epsilon = 1, 10^{-1}$, for $t=0.5$ and $N=32$. On the left, $\epsilon-$pseudospectra of the discretization matrix $\mathbf{L}$ via contour lines, for $t=0.5$ and $N=32$. }
    \label{fig:10}%
\end{figure}

Figure \ref{fig:10} provides two different visualizations of the pseudospectrum of the frozen coefficient matrix for a mesh of $32 \times 32$ nodes. In particular, the image on the right shows the contour lines of the pseudospectrum in the complex plane, while the image on the left shows an approximate picture of the pseudospectrum of $\boldsymbol{L}$, obtained by perturbing the semi-discretization matrix by random matrices $\boldsymbol{E}$ with norm $\|E\|= \epsilon.$ For these results, the entries $e_{ij}$ of the perturbation matrices were sampled from a normal complex distribution with mean $\langle e_{ij} \rangle=0$ and variance $\sigma^{2}=1$.\\

To approximate the distance between $\Lambda_{\epsilon}$ and the stability region of the method $S$ as a function of $\epsilon$ for different grid sizes, we proceed with the following steps.
$(1)$, we compute the pseudospectrum of $\boldsymbol{L}$ by the same procedure we used to obtain \figref{fig:10} (a); and 
$(2)$, we compute the distance $d(\Lambda_{\epsilon},S)$ as defined in Def.\ref{def:EpsilonStability}
\begin{equation*}
    \dist(\Lambda_{\epsilon}(\bsym{L}),S) := \max_{(\mu_{\epsilon},x)\in\Lambda_{\epsilon}(\bsym{L})\times S} \dist(\mu_{\epsilon},x).
\end{equation*}
\\

\begin{figure}[h]
    \centering
    \includegraphics[width=7.2cm]{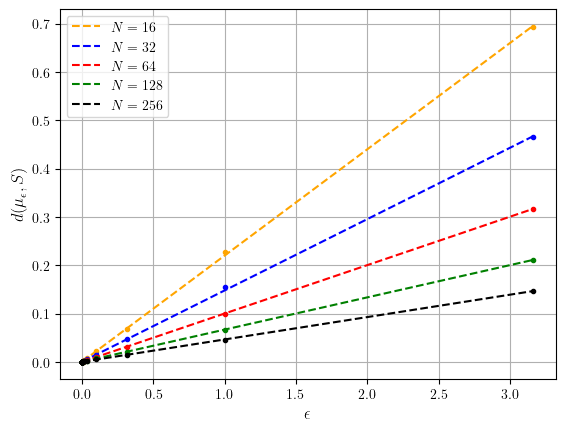}
    \caption{Distance of $\epsilon-$pseudospectrum to the Crank-Nicholson stability region as a function of $\epsilon$ for different grid sizes. Filled circles indicate measured distances, and the dashed line indicates the linear regression line. }
    \label{fig:5.13}
\end{figure}

Figure \ref{fig:5.13} shows that, for a fixed values of $N$ and $\Delta r$, the distance between the $\epsilon-$pseudospectrum and the stability region is a linear function of $\epsilon$, this is confirmed by the results of table \ref{Tab:Tcr2}. 
Moreover, we can see that the slope of the linear regression line depends on the grid size: bigger the angular resolution, smaller the slope. 
Notably, although these results depend on the number of random matrices considered and the method used for their generation, they also provide strong numerical evidence of the system being $\epsilon$-stable, for the frozen coefficient case.
Although this analysis is not sufficient to establish the global stability of the method, it provides insight into the overall behavior of the spectrum during the evolution.\\

\begin{table}[h]
\centering
\begin{tabular}{|cccc|}
\hline
\textbf{N} & \textbf{Slope}        & \textbf{Intercept}     & \textbf{R-value} \\ \hline
16         & $0.21987 \pm 0.00055$ & $0.0003 \pm 0.0003$    & 0.9999           \\ \hline
32         & $0.14757 \pm 0.00050$ & $0.0003 \pm 0.0003$    & 0.9998           \\ \hline
64         & $0.10014 \pm 0.00015$ & $-1.42e-6 \pm 0.00011$ & 0.9999           \\ \hline
128        & $0.06670 \pm 0.00007$ & $ 9.17e-5 \pm 5.46e-5$ & 0.9999           \\ \hline
256        & $0.04631 \pm 0.00006$ & $8.96e-5 \pm 4.60e-4$  & 0.9999           \\ \hline
\end{tabular}
 \caption{Best-fit parameters for the Linear Regressions for figure \ref{fig:5.13}}
\label{Tab:Tcr2}
\end{table}

\begin{figure}[h!]
    \centering
    \subfloat[\centering]
    {{\includegraphics[width=6.5cm]{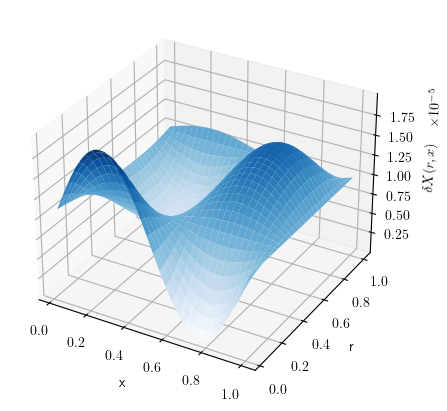} }}%
    \subfloat[\centering]
    {{\includegraphics[width=6.5cm]{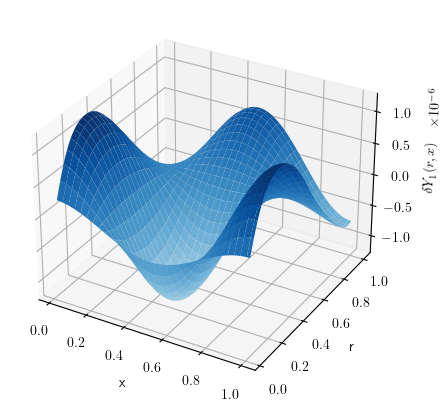} }}%
    \caption{Behavior of the numerical solution of the AHF equations with CN for Gowdy spacetime under initial condition modifications of the first component, for $N=64$ nodes, $\Delta r= 10^{-2}/N$ and $t=0.2$ \textbf{(a)} Solution for the first component $\delta X(r,x)$. \textbf{(b)} Solution for the second component $\delta Y_{1}(r,x)$.}
    \label{fig:Waterfallsgowdystable}%
\end{figure}

 Figure \ref{fig:Waterfallsgowdystable} shows both components of the numerical solution of the linearized AHF equations for an initial condition of the form of equation \ref{eq:InitialCondition}. We observe that contrary to the previous cases, in this case, no unstable oscillations appear as $r$ grows. In this case, the Hamiltonian and momentum constraints errors for the complete solution $(X+\delta X, Y_{1} + \delta Y_{1}, Y_{2} + \Delta Y_{2})$, are respectively,  $\|\mathcal{H} \| \approx 10^{-15}$ and $\|\mathcal{M} \| \approx 10^{-4}$, which still exceed the set tolerance levels. Nevertheless, in this case, it must be considered that the main source of error is caused by using a Fourier method for the constraints evaluation since it can be observed in figure \ref{fig:Waterfallsgowdystable} that the computed solution for the linearized AHF system $(\delta X, \delta Y_{1},\delta Y_{2})$ is not periodic with respect to the radial variable.\\

The previous analysis provides some insights regarding the behavior of the perturbed initial conditions in the AHF equations. Specifically, we demonstrated that the Fourier-based discretisation of the PDEs associated with each of the explored metrics induces numerical stability problems linked to the spectrum of the Fourier differentiation matrices. In particular, this problem manifested clearly for those systems that did not satisfy the hyperbolicity condition, such as FLRW, PFLRW and some specific cases of the Gowdy metric, depending on the variable $t$, which, in this case, is just a parameter.

\section{Construction of new initial data sets}
\label{section:new_id}

Despite the limitations of the AHF discussed in Secs.~\ref{section:test_error} and \ref{sec:Stability}, it remains possible to employ this formulation to potentially construct new initial data sets in cosmological spacetimes. 
In this section, we introduce two modifications to the hyperbolic system that allow us to obtain numerical solutions for specific cases.
We present numerical evidence of the viability of this modifications aiming to motivate future studies of this system in the context of cosmological spacetimes.\\

From the previous experiments, we observed that the tangential component of the mean three-dimensional extrinsic curvature of the two-dimensional hypersurfaces $S_r$, the field $Y_i$, has a strong impact on stability whenever it deviates from zero. The key idea behind the proposed modifications is to constrain this field in order to mitigate the observed instabilities.\\

\begin{figure}[h!]
    \centering
    \includegraphics[width=0.7\linewidth]{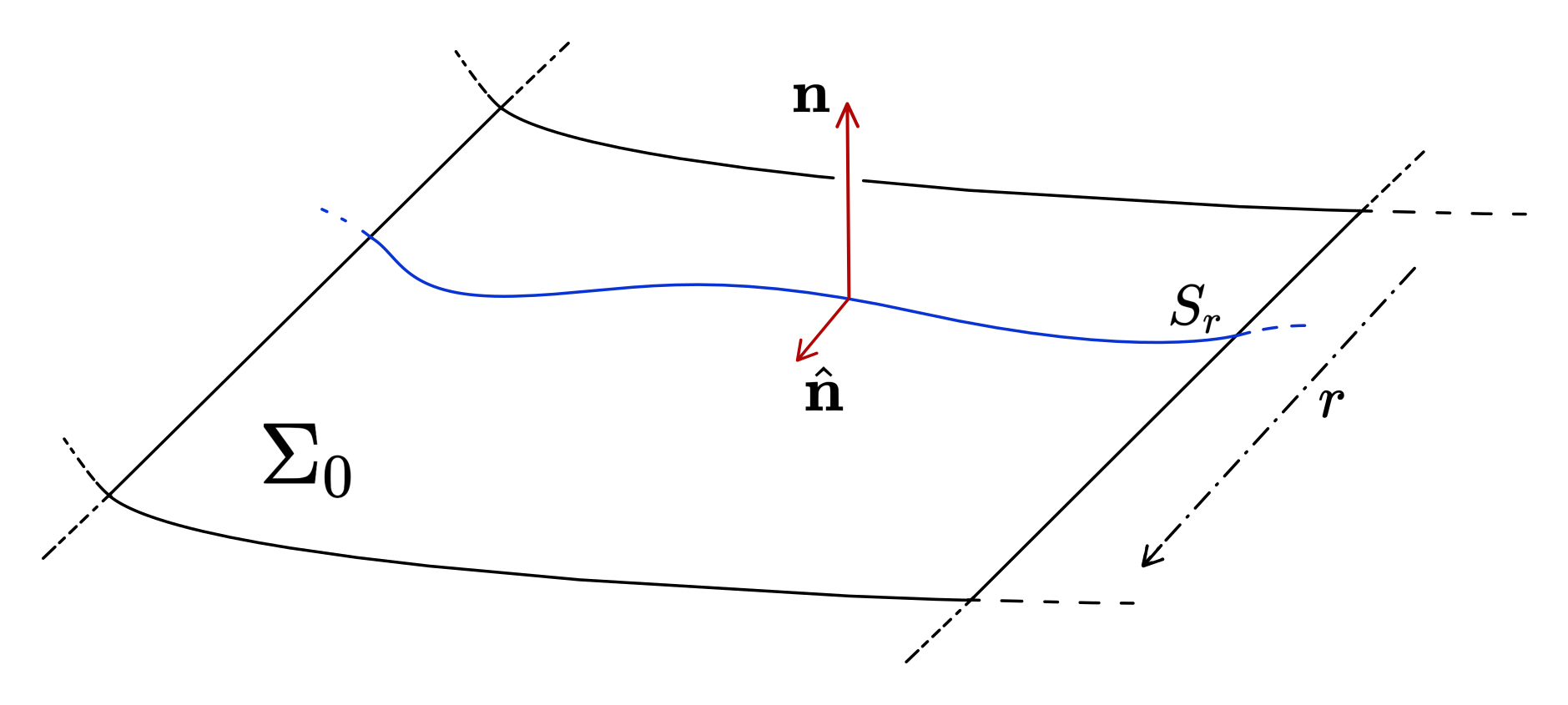}
    \caption{Illustration of the foliation of $\Sigma_0$ by the surfaces $S_r$ and their normal vectors, $\bsym{n}$ and $\hat{\bsym{n}}$ respectively. }
    \label{fig:ExtK_Illustration}
\end{figure}
% \begin{figure}[h!]
%     \centering
%     \includegraphics[width=0.75\linewidth]{figures/ExtK_illustration2.png}
%     % \caption{Enter Caption}
%     % \label{fig:placeholder}
% \end{figure}

To understand the implications of the constraint $Y_a=0$, we first recall the definition of the extrinsic curvature $K_{ab}$ (see for example \cite{Baumgarte} or \cite{Gourgoulhon}).
Let us denote by $\Sigma_0$ the initial spatial hypersurface of the foliation of the spacetime $M$ and by $n^a$ its normal vector. 
Then, the extrinsic curvature to $\Sigma_0$ can be computed as 
$$K_{ab} = \, \perp\joinrel\left(\nabla_a n_b\right),$$
where $\nabla$ is the covariant derivative in $M$ and $\perp\joinrel\left(\cdot\right)$ is the operator that project vectors from $TM$ to $T\Sigma_0$.
Therefore, the extrinsic curvature gives us a notion on the variation of the normal vector $n^a$ when moving on $\Sigma_0$.
On the other hand, as mentioned in Section~\ref{section:AHF_presentation}, the AHF recast the constraint equations after assuming that $\Sigma_0$ can be foliated by level surfaces of some \tit{global radius function} $r$. 
We have denoted this surfaces by $S_r$ and its normal vector as $\hat{n}^a$.
This \tit{second} foliation allows us to split every tensor field in $T\Sigma_0$ into its tangent and perpendicular parts to $TS_r$.  
In particular, $Y_a= h^a_{~b}\hat{n}^cK_{bc}$, as defined in eq.\eqref{ecs:formulae_of_decomposition}, represents the tangential component of extrinsic curvature.
% and is computed as $$Y_a = h^a_{~b}\hat{n}^cK_{bc}.$$
Therefore, restricting $Y_i$ is equivalent to impose conditions over the extrinsic curvature of $\Sigma_0$ in $M$, and then, the kind of foliations we can obtain for the spacetime $M$.
As consequence, the condition $Y_i=0$ for every point in $\Sigma_0$, when replaced in \eqref{drYeq}, implies that the relation
\begin{equation}
\label{Y0_Eq}
0 = \lapse \left( \frac{1}{2}\cd_iX + \cd_iZ - Z \ndot_i + \frac{1}{2}\ndot_iX + \ndot^j \ko_{ij} - \cd^j\ko_{ij} + 8 \pi \Jp_i \right), 
\end{equation}
must hold. 
Note that this constraint is direct consequence of Einstein equations and the geometric construction that allows for the AHF to be applicable (see Section~\ref{section:AHF_presentation} for details and \cite{racz2015constraints} for the original description).\\

In this section, we present two alternative approaches satisfy \eqref{Y0_Eq} in such a way that that can be used to numerically construct initial data sets.
In Section~\ref{sec:JparConstruction}, with the aim to show the viability of this approach, we solve it as an algebraic equation for $\Jp$.
Later, in Section~\ref{sec:H0Surfaces}, we explore another way to satisfy \eqref{Y0_Eq} while keeping the sources free by imposing conditions in the allowed foliations of $\Sigma_0$.\\

\subsection{Initial data sets of type 1}
\label{sec:JparConstruction}

In the simplest approach, \eqref{Y0_Eq} can be used to determine $J_i^{(||)}$. If the lapse function $\lapse$ is nonzero for each radial step, then $J_i^{(||)}$ can be computed from
\begin{equation*}
\Jp_i =  \frac{-1}{8\pi}\left( \frac{1}{2}\cd_iX + \cd_iZ - Z \ndot_i + \frac{1}{2}\ndot_iX + \ndot^j \ko_{ij} - \cd^j\ko_{ij}\right).
\end{equation*}

As a result, the hyperbolic system becomes
\begin{align}
\label{drXeq_Y0}
\pd{r}X & = \Ld{\shift}X + \lapse \left((Z - \frac{1}{2}X)H_j^{~j} - H_{ji}\ko^{ji} - 8 \pi \JT \right),\\
\label{Jp_Eq_Y0}
\Jp_i &=  \frac{-1}{8\pi}\left( \frac{1}{2}\cd_iX + \cd_iZ - Z \ndot_i + \frac{1}{2}\ndot_iX + \ndot^j \ko_{ij} - \cd^j\ko_{ij}\right),\\
\label{Zeq_Y0}
Z  &= \frac{1}{2X}\left( -\frac{1}{2}X^2 + \ko_{ij} \ko^{ij} - R + 16 \pi \rho \right).
\end{align}
This leads to a system consisting of one differential equation (eq.\eqref{drXeq_Y0}) and three algebraic equations (eq.\eqref{Jp_Eq_Y0} and eq.\eqref{Zeq_Y0}). 
It's clear that this is not an ideal case, as the standard way of constructing initial data sets is to leave the sources completely free. 
However, we point out that we are only dropping two of the four degrees of freedom in the energy sources. 
In particular, the energy density $\rho$ and the normal projection of the conserved current $\JT$ onto the hypersurfaces $S_{r}$ (the current in the direction $\hat{\bsym{n}}$) remain free for choice.
Sadly, study the situations in which this construction might be physically useful is out of the scope of this work and we hope to address it in the future.
For now, we use it to provide numerically evidence of the viability of this strategy.\\
% This still leaves us with a certain degree of freedom in the energy sources of the initial data set. \\

To illustrate the advantages of the above system of equations, let us consider the PFLRW metric eq. \eqref{SPFLRW_Metric} and the following initial condition for the field $X$
\begin{equation*}
  %\label{XIC_parametrization_Jpar}
  X^{(0)} = \bar{X}^{(0)} + \delta X_a,
\end{equation*}
where $\bar{X}^{(0)}$ is the analytic value corresponding to the PFLRW metric and  $\delta X_a$ for $a=1,2,3$, are some \tit{perturbations} that we chose as

\begin{align}
\label{deltaX0}
\delta X_0 &= 0,\\
\label{deltaX1}
\delta X_1 &= \phi_0 \left(\sin(\frac{\pi}{L}x_1) + \sin(\frac{\pi}{L}x_1) \right),\\
\label{deltaX2}
\delta X_2 &= \phi_0 \left( \cos\left(\frac{\pi}{L}x_1\right)\sin\left(\frac{\pi}{L}x_1\right) + \cos\left(\frac{\pi}{L}x_2\right)\sin\left(\frac{\pi}{L}x_2\right) \right),\\
\label{deltaX3}
\delta X_3 &= \phi_0 \left(\cos^2\left(\frac{\pi}{L}x_1\right) + \cos^2\left(\frac{\pi}{L}x_2\right) \right),
\end{align}
with $\phi_0 = 10^{-8}$ the parameter controlling the perturbation in the initial condition.\\

We compute the sources by evaluating the original constraint equations, eqs. (\ref{eq:HC_operator}) and (\ref{eq:MC_operator}) and solving for $\rho$ and $J_i$. When the PFLRW metric is specified as the, the energy density $\rho$ and the three-dimensional current $J_i$ become functions of $\phi(r,x_1,x_2)$. Applying the $2+1$ decomposition to $J_i$, we obtain $J^{(\perp)}$, which, along $\rho$, are used to close the system. $J^{(||)}_i$ is discarded and recomputed from eq. (\ref{Jp_Eq_Y0}) at each radial step.\\

\begin{figure}[ht] 
\includegraphics[width = 0.8\textwidth,center]{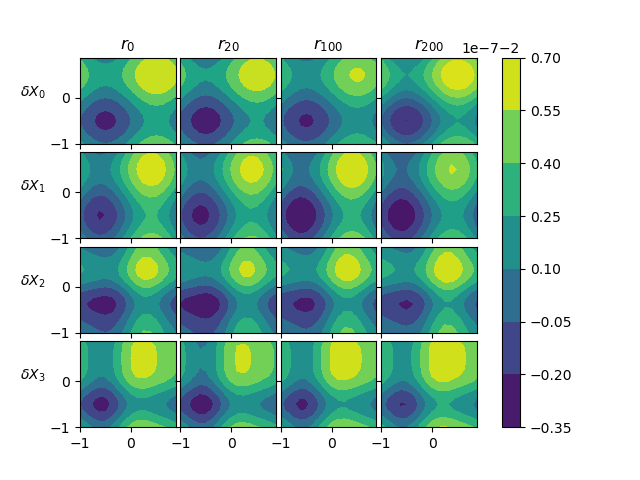}
\caption{Snapshots of the solution of $X(r,x_1,x_2)$, $r=r_k = -L + k\,\Delta r$, for different initial conditions. The first row show $x_1-x_2$ slices of the solutions for the original initial condition, \ie, without modification. The other three rows show the deviation $\delta X = X - X_{background}$ solution for the modifications of the initial conditions presented in \cref{deltaX1,deltaX2,deltaX3}. The color bar represents the values of $X$ in the scale of $X - X_{background}$ where $X_{background}$ is the unperturbed solution. These results where achieved for $N = 16$, $Factor = 16$, and no filter was used.}
\label{RHC_ICVariation_PFLRW}
\end{figure}

In figure \ref{RHC_ICVariation_PFLRW}, we display some contour plots of the numerical solutions obtained from the different initial conditions eqs. (\ref{deltaX1}-\ref{deltaX3}). By evaluating the constraint equations (eqs. (\ref{eq:HC_operator}) and (\ref{eq:MC_operator})) from these numerical solutions, we obtain a constraint violation value around $10^{-13}$ with $N = 16$ and $Factor = 16$, indicating that we are indeed obtaining solutions to the constraint equations. It is worth noticing that the errors do not significantly improve by increasing the grid, the number of radial steps, or changing the filter. \\

This experiment allows us to conclude that it is possible to obtain new initial data sets with this approach at least at the perturbation regime. Further study is needed to determine the implications of the assumption $Y_i=0$ and how physic is the constraint $\Jp_i$. 
Additionally, in this experiment $\bsym{\shift}$ is zero, in consequence, \eqref{drXeq_Y0} becomes an ODE. The derivatives of $X$ are only present in the construction of $\Jp_i$ making this a very simplified test case.

\subsection{Initial data sets of type 2}
\label{sec:H0Surfaces}

We conclude this section by exploring a different way to solve \cref{Y0_Eq} without losing generality in the degrees of freedom of the sources, that is, keeping $J_i^{(||)}$ as a free function. 
The idea for this approach arises from the following observation: if we assume $\beta_i=0$, equation \cref{drXeq_Y0} do not depend anymore on angular derivatives of $X$. Thus, by assuming $H^i_{~i}=0$, indicating that we are foliating $\Sigma_0$ with minimal surfaces, the equation \cref{drXeq_Y0} becomes independent of $X$. Therefore, it can be solved through integration along the radial coordinate by
\begin{equation}\label{Xsol}
X(r,x_1,x_2) = \int_{r_0}^{r}\lapse(H_{ji}\ko^{ji} - 8 \pi \JT) dr + F(x_1,x_2),
\end{equation}
with $F(x_1,x_2)$ being an arbitrary function that depends only on the angular variables. 
Since $H_{ji}$, $\kappa^{ji}$, and $\JT$ are all known functions, the integral term in eq. (\ref{Xsol}) can be computed analytically or numerically using conventional methods. Therefore, to determine $X$, we have to provide a way to compute the function $F(x_1,x_2)$ on the initial surface $S_{r_{0}}$. To achieve this, we solve eq. (\ref{Y0_Eq}) for $\cd_iX$, leading to
\begin{equation*}
\cd_iX = \frac{-1}{\frac{1}{4} + \frac{Z_0}{2X}} \left( - \frac{\cd_iZ_0}{2X} - Z \ndot_i + \frac{1}{2}\ndot_iX + \ndot^j \ko_{ij} - \cd^j\ko_{ij} + 8 \pi \Jp_i \right),
\end{equation*}  
where $Z_0 = R -\ko_{ij} \ko^{ij} - 16 \pi \rho$.
Then, substituting $X = I + F$, where $I$ is the integral term of \cref{Xsol}, we obtain 
\begin{equation}
\label{eq:Y0_F_equation}
  \cd_i F  = \hat{G}_i(X) - \cd_i I = G_i(F;r,x_1,x_2), 
\end{equation}
with
\begin{equation*}
    \hat{G}_i(X)  := \frac{-1}{\frac{1}{4} + \frac{Z_0}{2X}} \left( - \frac{\cd_iZ_0}{2X} - Z \ndot_i + \frac{1}{2}\ndot_iX + \ndot^j \ko_{ij} - \cd^j\ko_{ij} + 8 \pi \Jp_i \right).
\end{equation*}
Finally, applying covariant derivatives on both sides of the equation and contracting with the angular metric $h_{ab}$, we obtain the following second order PDE for $F$
\begin{equation}
\label{EllipticFEq}
\Delta F = \cd^i\cd_i F = \cd^iG_i = g\left(F,\pd{}F;r,x_1,x_2\right).
\end{equation}
 
Because  the principal operator of this equation is the Laplacian on $S_{r}$ for some fixed value of $r$, it can be solved by standard numerical methods for elliptic PDEs, such as the fixed point or Newton algorithms. However, these methods require a further analytical analysis in order to guarantee convergences of the solution. Furthermore, since the discrete Laplacian obtained through Fourier differentiation does not lead to an invertible matrix in the angular grid (see for instance \cite{Trefethen} or \cite{Kopriva}), the application of these methods is not a trivial task. Therefore, to avoid such complications, we simply transform the elliptic problem \cref{EllipticFEq} into the following parabolic initial value problem
\begin{align} \label{ParabolicFEq}
\pd{t}F   &= \Delta F - g\left(F,\pd{}F;r,x_1,x_2\right),\\
F|_{t_{0}}&= F_0 \,
\end{align}
where $t$ is a free parameter and $F_0$ is a known function on $S_{r}$. Local existence and uniqueness of solutions for this kind of parabolic equations are well established. See, for instance, \cite{TaylorPDEIII}, ch. 15. In particular, it is known that given initial condition of $F$ belonging to $C^1(M)$, the solution of \cref{ParabolicFEq} should exists and is unique on $(t,\vec{x})\in[0,T]\times M$ for some $T>0$, where $M$ is a Riemannian manifold.\\

At this point, it is worth discussing the type of initial data that we can obtain from this approach. We recall that in addition to imposing the condition $Y_i=0$, we also assume $\beta_i=0$ and $H^i_{~i}=0$, which are restrictions on the foliations $S_r$ along the coordinate $r$. The first condition simply means that the normal vector to the surfaces $S_r$ is aligned with the coordinate $r$, and the second condition requires that the surfaces $S_r$ are minimal for any fixed value of $r$. One way to easily achieve this condition is by demanding that the components of the spatial metric $\gamma_{ab}$ are independent of $r$, which implies that the metric is $\mathbb{S}^1$ symmetric. However, this does not necessarily mean that the resulting initial data set $(\gamma_{ab},K_{ab},\rho,J_{a})$ will be $\mathbb{S}^1$ symmetric, as the components of the extrinsic curvature can still depend on the $r$-coordinates through eq. (\ref{Xsol}).\\

It is important to note that eq. (\ref{eq:Y0_F_equation}) is equivalent to the original $Y_i$ equation eq. (\ref{drYeq}) under the conditions $Y_i = \shift_i = 0$ and $H^{i}_{~i}=0$. Therefore, the quantity
\begin{equation}
\label{eq:Parabolic_Residual}
R = ||\cd{i}F + \cd{i}I - \hat{G}_i(X)||,
\end{equation}
can be used as a measure of the \textit{fulfillment} of the hyperbolic constraint equations obtained through the solution of eq. (\ref{ParabolicFEq}). Therefore, we can state two stopping criteria: $(1)$ Since the solutions of eq. (\ref{ParabolicFEq}) must reach a steady state, if $X_n$ and $X_{n-1}$ are solutions obtained with successive $t$ steps $t_n$ and $t_{n-1}$, we expect $||X_n - X_{n+1}||$ to converge to zero. Then, when $||X_n - X_{n+1}||$ gets under a certain tolerance, we can stop the evolution. $(2)$ Since eq. (\ref{eq:Parabolic_Residual}) tells us if the equation for $Y_i$ is being fulfilled, when $R$ gets under a certain tolerance, we can stop the evolution. The difference between these two stopping criteria is displayed below for a particular test case.\\ 

With the above in mind, as in the previous sections, we need to close the system by setting the free fields. To do so we will consider the following $4-$dimensional metric
\begin{equation}
\label{Met_ParabExample}
  g_{\mu\nu} = \begin{pmatrix}
                -f(t,x_1) & 0        & 0        & 0        \\
                0         & g(t,x_1) & 0        & 0        \\
                0         & 0        & g(t,x_1) & 0        \\
                0         & 0        & 0        & g(t,x_1) \\
              \end{pmatrix},
\end
{equation}
with
\begin{equation*}
     f(t,x_1) = g(t,x_1) = t^2\left(1-\frac{1}{4}\cos\left(\frac{\pi}{L}x_1\right)\right).\\
\end{equation*}  

By $3+1$ decomposition we can compute $\gamma_{ab}$ and the analytical $K_{ab}$. Then through the $2+1$ decomposition, we can obtain the geometrical free fields, finding that $H_{ij}=\mathring{k}_{ij}=0$. Additionally, by evaluating the constraint equations, eqs. (\ref{eq:HC_operator}) and (\ref{eq:MC_operator}), we can compute the sources $\rho$ and $J_i$ as functions of $g(t,x_1)$, obtaining $\JT=0$ and $\Jp\neq0$.
Therefore, with the conditions $H_{ij}=\mathring{k}_{ij}=0$ and $\JT=0$, we can drop the integral in (\ref{Xsol}) to obtain $X=F$.\\

\begin{figure}[h!]
\begin{subfigure}{0.51\textwidth}
  \includegraphics[width = 1.05\textwidth,left]{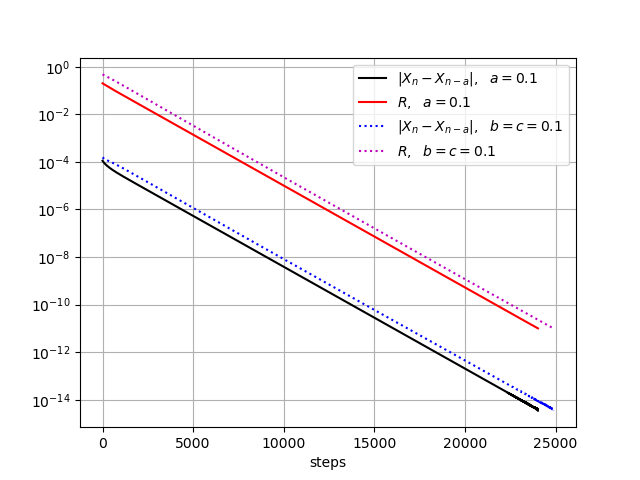}
  \caption{}
\end{subfigure}
\begin{subfigure}{0.51\textwidth}
  \includegraphics[width = 1.05\textwidth,right]{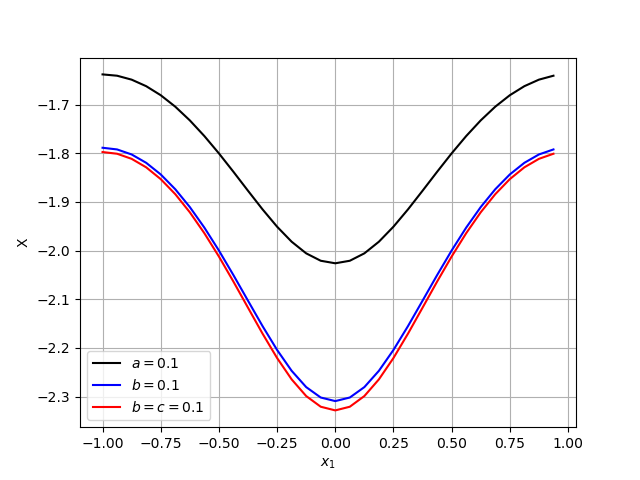}
  \caption{}
\end{subfigure}
\caption{Plots achieved in a grid of $32$ nodes, RK4 as the integrator and a \tit{temporal} size-step of $10^{-4}$. \tbf{(a)} Parabolic relaxation of the solution.  \tbf{(b)} Plot of the resulting $X$ once $R<10^{-11}$ for different initial conditions.
The parameters $a,~b$ and $c$ correspond to an initial condition parametrization of $F$ as in \cref{FIC_parametrization}.} 
\label{ParabExample_ICmod}
\end{figure}

\begin{figure}[h!]
\begin{subfigure}{0.51\textwidth}
  \includegraphics[width = 1.05\textwidth,left]{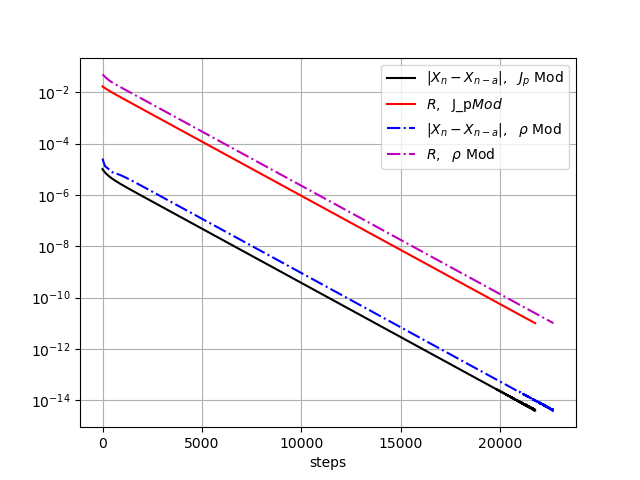}
  \caption{}
\end{subfigure}
\begin{subfigure}{0.51\textwidth}
  \includegraphics[width = 1.05\textwidth,right]{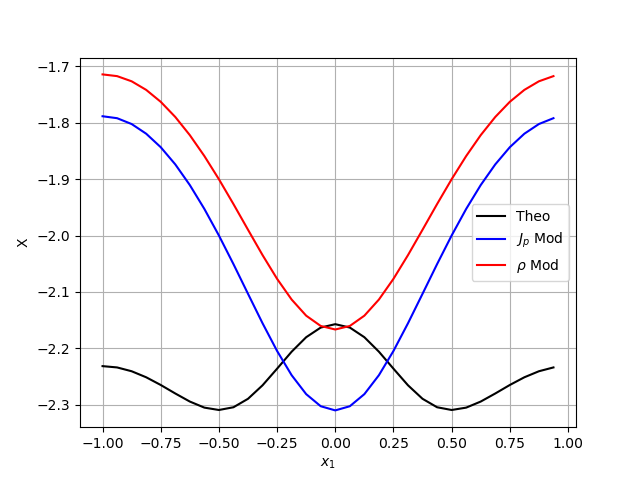}
  \caption{}
\end{subfigure}
\caption{Plots were achieved in a grid of $32$ nodes, RK4 as the integrator and a radial step-size of $10^{-4}$. \tbf{(a)} Both, the difference of consecutive steps and the residual $R$ evaluated at these steps converge to zero. \tbf{(b)} Plot of the resulting $X$, once $R<10^{-11}$, when the sources are modified by multiplying them by $1-\varepsilon$ with $\varepsilon = 0.001$.}
\label{ParabExample_SourceMod}
\end{figure} 

To illustrate the behavior of this approach, we perform two kinds of experiments: initial condition and source modifications. The initial condition modifications take the form 
\begin{equation}
    \label{FIC_parametrization}
    F^{(0)} = \bar{F}^{(0)} + a  + b\sin\left(\frac{\pi}{L}x_1\right) +c\cos\left(\frac{\pi}{L}x_1\right).
\end{equation}
applied to $F$ where $\bar{F}^{(0)}$ is computed from the analytical expression obtained from the $3+1$ and $2+1$ decompositions of eq. (\ref{Met_ParabExample}). 
In \figref{ParabExample_ICmod}\tbf{(a)}), we can see how the difference between consecutive RK4 steps and the residual $R$ evaluated at these steps converge to zero. This behavior displays the convergence to a \tit{steady state} of the solutions of \cref{ParabolicFEq}.
It is interesting to note that different choices of the initial condition for \cref{ParabolicFEq} produce different solutions (see \figref{ParabExample_ICmod}\tbf{(b)}). The constraint violation associated to these solutions is in the order of $10^{-12}$; therefore, we are producing new initial data sets from the different initial conditions.\\

For the sources modifications, we multiply the source function by a factor $1-\varepsilon$ to alter the amplitude of the source affecting the whole behavior of the solution. 
This simple source modification produces solutions with a constraint violation around $10^{-12}$ (see \figref{ParabExample_SourceMod}). By these experiments, the parabolic relaxation method has probed to be a promising tool to find solutions for the AHF under the condition $Y_i=0$. However, we have noticed that more drastic modifications can produce solutions that do not fulfill the constraint equations. Thus, more investigation into this approach is needed to evaluate the origin of these limitations.

\section{Discussion and conclusions}\label{section:discussion}

In this work, we explored the Algebraic–Hyperbolic Formulation (AHF) of the Einstein constraint equations in a cosmological context.
Specifically, we applied the AHF to cosmological models with $\mathbb{T}^3$ spatial topology, focusing on two numerically and physically relevant cases: $\mathbb{T}^3$–Gowdy spacetimes and perturbed FLRW universes. Within this framework, we implemented the AHF numerically using a pseudo-spectral Fourier algorithm.
Although the method successfully reproduced known analytical solutions for both spacetimes, we found numerical evidence of significant stability issues that could not be mitigated by increasing the resolution or applying standard anti-aliasing techniques. In particular, for the PFLRW metric (Fig.~\ref{fig:PFLRWTest}), the convergence test failed.\\

These findings motivated the stability analysis presented in Section~\ref{sec:Stability}.
In Theorem~\ref{teorema}, we showed that, for any metric close to FLRW, this instability is unavoidable.
We realized that the instability of the AHF in this case is governed by the spectral properties of the Fourier differentiation matrix, which, together with the block structure of the semi-discretization matrix $\mathbf{L}$, allowed us to conclude that its eigenvalue distribution lay near the positive real axis, thus falling outside the stability region of any RK method.
It is worth mentioning that, for this pessimistic result to hold, the AHF system has to be applied to a \emph{FLRW-like} metric using a Fourier method of lines. 
Changing any of these ingredients leads us to different conclusions.\\

% \CC{corrected some inconsistencies in this conclusion, the spectrum of the Fourier matrix lie in the imaginary axis}

To test this in a different spacetime, we examined the stability for the Gowdy metric.
In this case, the parameter $t$ plays a central role in the stability of the system due to its connection with the hyperbolicity condition (see Eq.~\eqref{ec:hyperboliccondition} and the related discussion in Section~\ref{section:AHF_presentation}). We found that the numerical scheme remains stable when the solutions satisfy the hyperbolicity condition for all values of the radial or foliation variable. Additionally, we observed that stable solutions need not be periodic along the foliation coordinate. However, since our spatial differentiation scheme assumes $\mathbb{T}^3$-periodicity, our current implementation cannot guarantee that the constraint equations are fully satisfied in such cases. This limitation could potentially be addressed by adopting a more flexible differentiation scheme, at least during the ADM constraint evaluation.\\

In Section~\ref{section:new_id}, we provide numerical evidence of the viability of the AHF and the Fourier-based MoL when further geometrical restrictions are applied.  
We explored two possible modifications of the original AHF that mitigate its unstable behavior. Our experiments showed that when the tangential projection of the mean three-dimensional extrinsic curvature onto the two-dimensional foliations is nonzero, the hyperbolic constraints become numerically unstable, significantly degrading the accuracy of the resulting initial data. Therefore, we proposed two distinct strategies to address this issue and construct new initial data sets. Although still preliminary, the experiments presented in Sections~\ref{sec:JparConstruction} and \ref{sec:H0Surfaces} demonstrate two promising modifications of the original AHF that enable the numerical construction of periodic initial data sets in cosmological contexts.\\

The first approach enables the numerical construction of initial data sets in which one component of the projected current onto the two-surface is determined by the geometry, while the remaining components are treated as freely specifiable functions. This restriction naturally limits the range of admissible energy distributions and, consequently, the class of physical systems that can be modeled with such initial data.
The question of which physical systems can benefit from this kind of description remains open, and we expect to address it in the future.
In contrast, the second approach yields initial data with entirely free energy quantities by imposing a constraint on the mean curvature of the two-dimensional foliations.
In particular, it assumes that the initial spatial hypersurface can be foliated by maximal surfaces with vanishing shift. 
Identifying which class of three dimensional hypersurfaces fulfills these conditions is another unresolved question that we intend to tackle in future works.\\

While the proposed modifications show encouraging behavior, further theoretical and numerical analysis is required to fully assess their robustness and physical significance. Nevertheless, the results presented here highlight the potential of hyperbolic formulations as viable alternatives to traditional elliptic approaches in numerical cosmology, opening the way to new strategies for constructing initial data sets.

\section*{Acknowledgments}
%This work was supported by Patrimonio Autonomo-Fondo Nacional de Financiamiento para la Ciencia, la Tecnología y la Innovación Francisco José de Caldas (MIN-CIENCIAS-COLOMBIA) Grant No. 110685269447 RC-80740-465-2020, and the Vicerectoria de Investigaciones of the Universidad del Valle in the project 71254: Dark universe and Large Scale Structure. 

This work was supported by Patrimonio Autónomo - Fondo Nacional de Financiamiento para la Ciencia, la Tecnología y la Innovación Francisco José de Caldas (MINCIENCIAS - COLOMBIA) Grant No. 110685269447 RC-80740-465-202, projects 69723 and 69553.

\bibliographystyle{unsrt}
\bibliography{references}

@book{Butcher2016,
   author = {J. C. Butcher},
   doi = {10.1002/9781119121534},
   isbn = {9781119121503},
   month = {7},
   publisher = {Wiley},
   title = {Numerical Methods for Ordinary Differential Equations},
   url = {https://onlinelibrary.wiley.com/doi/book/10.1002/9781119121534},
   year = {2016}
}

@article{Aurrekoetxea2022,
  author = {Aurrekoetxea, Josu C. and Clough, Katy and Lim, E.A.},
  title = {CTTK: A new method to solve the initial data constraints in numerical relativity},
  journal = {arXiv preprint arXiv:2207.03125},
  year = {2022}
}

@article{Macpherson2019,
  title={Numerical relativity for cosmology},
  author={Macpherson, Hayley J. and Price, Daniel J. and Lasky, Paul D.},
  journal={Physical Review D},
  volume={99},
  number={6},
  pages={063522},
  year={2019},
  doi={10.1103/PhysRevD.99.063522}
}

@article{Bentivegna2024,
  title={Numerical Cosmology: Modelling Inhomogeneities in General Relativity},
  author={Bentivegna, Eloisa and Korzy{\'n}ski, Miko{\l}aj},
  journal={Living Reviews in Relativity},
  year={2024},
  doi={10.1007/s41114-024-00046-0}
}

@article{Amorim2009,
  title={Numerical evolution of polarized Gowdy $T^3$ cosmological models},
  author={Amorim, R. and Rebouças, M. J. and Gomero, G. I.},
  journal={Classical and Quantum Gravity},
  volume={26},
  number={3},
  pages={035018},
  year={2009},
  publisher={IOP Publishing},
  doi={10.1088/0264-9381/26/3/035018}
}

@article{Gambini2005,
  title={Quantum evolution of the Gowdy universe},
  author={Gambini, Rodolfo and Pullin, Jorge},
  journal={Classical and Quantum Gravity},
  volume={22},
  number={17},
  pages={S479--S488},
  year={2005},
  publisher={IOP Publishing},
  doi={10.1088/0264-9381/22/17/S14}
}

@article{aurrekoetxea2025cosmology,
  title={Cosmology using numerical relativity: JC Aurrekoetxea et al.},
  author={Aurrekoetxea, Josu C and Clough, Katy and Lim, Eugene A},
  journal={Living Reviews in Relativity},
  volume={28},
  number={1},
  pages={5},
  year={2025},
  publisher={Springer}
}

@book{do1992riemannian,
  title={Riemannian geometry},
  author={Do Carmo, Manfredo Perdigao and Flaherty Francis, J},
  volume={6},
  year={1992},
  publisher={Springer}
}

@article{racz2014cauchy,
	title={Cauchy problem as a two-surface based ‘geometrodynamics’},
	author={R{\'a}cz, Istv{\'a}n},
	journal={Classical and Quantum Gravity},
	volume={32},
	number={1},
	pages={015006},
	year={2014},
	publisher={IOP Publishing}
}

@article{racz2014bianchi,
  title={Is the Bianchi identity always hyperbolic?},
  author={R{\'a}cz, Istv{\'a}n},
  journal={Classical and Quantum Gravity},
  volume={31},
  number={15},
  pages={155004},
  year={2014},
  publisher={IOP Publishing}
}

@article{racz2015constraints,
  title={Constraints as evolutionary systems},
  author={R{\'a}cz, Istv{\'a}n},
  journal={Classical and Quantum Gravity},
  volume={33},
  number={1},
  pages={015014},
  year={2015},
  publisher={IOP Publishing}
}

@article{Ma_Berts,
title = {Cosmological perturbation theory in the synchronous and conformal Newtonian gauges},
author = {Ma, C and Bertschinger, E},
journal = {Astrophysical Journal},
number = 1,
volume = 455,
place = {United States},
year = {1995},
month = {12}
}

@book{Peebles,
  title={The large-scale structure of the universe},
  author={Peebles, Phillip James Edwin},
  volume={98},
  year={2020},
  publisher={Princeton university press}
}

@book{Dodelson_2ed,
  title={Modern cosmology},
  author={Dodelson, Scott and Schmidt, Fabian},
  year={2020},
  publisher={Academic press}
}

@article{ringstrom2010cosmic,
  title={Cosmic censorship for Gowdy spacetimes},
  author={Ringstr{\"o}m, Hans},
  journal={Living Reviews in Relativity},
  volume={13},
  pages={1--59},
  year={2010},
  publisher={Springer}
}

@book{Wald,
  author = "Wald, Robert M.",
  title = "{General Relativity}",
  doi = "10.7208/chicago/9780226870373.001.0001",
  publisher = "Chicago Univ. Pr.",
  address = "Chicago, USA",
  year = "1984"
}

@book{DurrerCMB,
  author= "Ruth Durrer",
  title = "The Cosmic Microwave Background",
  year  = "2008",
  publisher = "Cambridge University Press"
}

@article{Australians2017,
  title = {Inhomogeneous cosmology with numerical relativity},
  author = {Macpherson, Hayley J. and Lasky, Paul D. and Price, Daniel J.},
  journal = {Phys. Rev. D},
  volume = {95},
  issue = {6},
  pages = {064028},
  numpages = {13},
  year = {2017},
  month = {Mar},
  publisher = {American Physical Society},
  doi = {10.1103/PhysRevD.95.064028},
  url = {https://link.aps.org/doi/10.1103/PhysRevD.95.064028}
}

@article{Australians2019,
  title = {Einstein's Universe: Cosmological structure formation in numerical relativity},
  author = {Macpherson, Hayley J. and Price, Daniel J. and Lasky, Paul D.},
  journal = {Phys. Rev. D},
  volume = {99},
  issue = {6},
  pages = {063522},
  numpages = {18},
  year = {2019},
  month = {Mar},
  publisher = {American Physical Society},
  doi = {10.1103/PhysRevD.99.063522},
  url = {https://link.aps.org/doi/10.1103/PhysRevD.99.063522}
}

@book{Baumgarte,
  title={Numerical relativity: solving Einstein's equations on the computer},
  author={Baumgarte, Thomas W and Shapiro, Stuart L},
  year={2010},
  publisher={Cambridge University Press}
}

@book{alcubierre2008introduction,
  title={Introduction to 3+ 1 numerical relativity},
  author={Alcubierre, Miguel},
  volume={140},
  year={2008},
  publisher={OUP Oxford}
}

@book{Gourgoulhon,
author= "Eric Gourgoulhon",
title = "3+1 Formalism and Bases of Numerical Relativity",
year  = "2007",
publisher = "",
address = "https://doi.org/10.48550/arXiv.gr-qc/0703035"
}

@article{Alcubierre_Testbeds,
	doi = {10.1088/0264-9381/21/2/019},
	url = {https://doi.org/10.1088/0264-9381/21/2/019},
	year = 2003,
	month = {dec},
	publisher = {{IOP} Publishing},
	volume = {21},
	number = {2},
	pages = {589--613},
	author = {Miguel Alcubierre and Gabrielle Allen and Carles Bona and David Fiske and Tom Goodale and F Siddhartha Guzm{\'{a}}n and Ian Hawke and Scott H Hawley and Sascha Husa and Michael Koppitz and Christiane Lechner and Denis Pollney and David Rideout and Marcelo Salgado and Erik Schnetter and Edward Seidel and Hisa-aki Shinkai and Deirdre Shoemaker and B{\'{e}}la Szil{\'{a}}gyi and Ryoji Takahashi and Jeff Winicour},
	title = {Towards standard testbeds for numerical relativity},
	journal = {Classical and Quantum Gravity}
}

@article{Garfinkle2020,
  title = {Cosmological initial data for numerical relativity},
  author = {Garfinkle, David and Mead, Lawrence},
  journal = {Phys. Rev. D},
  volume = {102},
  issue = {4},
  pages = {044022},
  numpages = {9},
  year = {2020},
  month = {Aug},
  publisher = {American Physical Society},
  doi = {10.1103/PhysRevD.102.044022},
  url = {https://link.aps.org/doi/10.1103/PhysRevD.102.044022}
}

@book{TaylorPDEIII,
  title={Partial differential equations II: Qualitative studies of linear equations},
  author={Taylor, Michael},
  volume={116},
  year={2013},
  publisher={Springer Science \& Business Media}
}

@book{Trefethen,
author= "Trefethen, Lloyd N",
title = "Spectral methods in MATLAB",
year  = "2000",
publisher = "SIAM",
DOI = {https://doi.org/10.1137/1.9780898719598},
address = "https://doi.org/10.1137/1.9780898719598"
}

@book{Kopriva,
  title={Implementing spectral methods for partial differential equations: Algorithms for scientists and engineers},
  author={Kopriva, David A},
  year={2009},
  publisher={Springer Science \& Business Media}
}

@book{Canuto,
  title={Spectral methods: evolution to complex geometries and applications to fluid dynamics},
  author={Canuto, Claudio and Hussaini, M Yousuff and Quarteroni, Alfio and Zang, Thomas A},
  year={2007},
  publisher={Springer Science \& Business Media}
}

@book{Stoer_Bulirsch,
  title={Introduction to numerical analysis},
  author={Stoer, Josef and Bulirsch, Roland and Bartels, R and Gautschi, Walter and Witzgall, Christoph},
  volume={1993},
  year={1980},
  publisher={Springer}
}

@article{higham1993stiffness,
  title={Stiffness of odes},
  author={Higham, Desmond J and Trefethen, Lloyd N},
  journal={BIT Numerical Mathematics},
  volume={33},
  pages={285--303},
  year={1993},
  publisher={Springer}
}

@book{trefethen2005spectra,
  title={Spectra and Pseudospectra: The Behavior of Nonnormal Matrices and Operators},
  author={Trefethen, Lloyd N and Embree, Mark},
  year={2005},
  publisher={Princeton University Press}
}

@article{mortonlax,
  title={Lax-stability vs. eigenvalue stability of spectral methods},
  author={Morton, From KW and MJ, Baines}
}

@article{reddy1992stability,
  title={Stability of the method of lines},
  author={Reddy, Satish C and Trefethen, Lloyd N},
  journal={Numerische Mathematik},
  volume={62},
  number={1},
  pages={235--267},
  year={1992},
  publisher={Springer}
}

% \newpage
\appendix

\section{$3$-dimensional constraints evaluation}
\label{appendix:Constriant_Checker}

    \subsection{Additional test cases}
    In addition to the Gowdy and PFLRW metrics presented in Section~\ref{sec:exact_solutions}, we introduce to gauge-wave metrics to test our implementation of the $3$-dimensional constraint equations  
    
    \begin{enumerate}
        \item \tbf{The Minkowski gauge-wave metric (MXY):} Even tough this spacetime does not have a global $\mbb{T}^3$ spatial topology, we can assume that it is composed by an infinite union of copies of a cube with $\mbb{T}^3$ spatial topology globally parameterized by periodic-coordinates $(r,x_1,x_2)$. Thus, following \cite{Alcubierre_Testbeds}, we apply a coordinate transformation so the resulting metric is not trivial. 
    Let us denote by $\{t',r',x_1',x_2'\}$ the usual Cartesian coordinates such that the Minkowski metric takes its usual form $g_{\mu\nu}' = \eta_{\mu\nu} = \text{Diag}(-1,1,1,1)$\footnote{The prime in $g_{\mu\nu}'$ indicates that it is the metric computed in the primed coordinates $\{t',r',x_1',x_2'\}$}.
    Now, we define a new set of coordinates $\{\hat{t},\hat{r},\hat{x}_1,\hat{x}_2\}$ related to the prime coordinates by 
    \begin{align*}
        \{\hat{t},\hat{r},\hat{x}_1,\hat{x}_2\} &\leftarrow \{t'+\frac{Ad}{2\pi}\cos\left(\frac{2\pi}{d}(r'-t')\right),r'-\frac{Ad}{2\pi}\cos\left(\frac{2\pi}{d}(r'-t')\right),x_1',x_2'\}, 
    \end{align*}
    where $A$ and $d$ are free parameters of the transformation.
    
    If $\hat{J}$ represents the Jacobian of this transformation, the metric in the new coordinates $\hat{g}_{\mu\nu}$ can be computed as 
    $$\hat{g}^{\mu\nu} = (\hat{J}^T\,g'\,\hat{J})^{\mu\nu}.$$
    
    Additionally, we take a second coordinate transformation given by 
    \begin{align*}
        \{t,r,x_1,x_2\} &\leftarrow \{\hat{t},\frac{1}{\sqrt{2}}(\hat{r}-\hat{x}_1),\frac{1}{\sqrt{2}}(\hat{r}+\hat{x}_1),\hat{x}_2\}, 
    \end{align*}
    which, if $J$ denotes its Jacobian, leads the Minkowski metric to the following form 
    \begin{equation}
    \label{MXY_Metric}
    g_{\mu\nu} = \left((J^T\,\hat{J}^T\,\eta\,\hat{J}\,J)^{-1}\right)_{\mu\nu} = 
    \begin{pmatrix}
      -(1-M) & 0 & 0 & 0\\
       0 & 1-\frac{1}{2}M & \frac{1}{2}M & 0\\
       0 & \frac{1}{2}M & 1-\frac{1}{2}M & 0\\
       0 & 0 & 0 & 1\\
    \end{pmatrix},
    \end{equation}
    where $M = M(t,r,x_1) = A\sin\left(\frac{\pi(2t + \sqrt{2}(x_1-r))}{d}\right)$.  
    
    \item \tbf{Gowdy gauge wave (GRX):} Since the Gowdy metric in standard coordinates depends only on one spatial variable, it does not offer a good test case for our numerical implementation. Therefore, in analogy to the previous metric, we modify it through the following coordinate transformation. 
    Denoting by $\{t',r',x_1',x_2'\}$ the usual Cartesian coordinates that give place to the metric $g_{\mu\nu}'$ as written in eq. \eqref{Gowdy_Metric}, we define the coordinate transformation
    
    \begin{equation*}
    \label{Gowdy_Rotation}
    \{t,r,x_1, x_2\} \leftarrow \{t',\frac{1}{\sqrt{2}}(r'-x'_1),\frac{1}{\sqrt{2}}(r'+x'_1),x'_2\}. 
    \end{equation*}
    
    Note that this coordinate transformation might be interpreted as a rotation around $x_2$ axis by an angle $\theta = \dfrac{\pi}{4}$.
    
    In these new coordinates, the Gowdy $\mbb{T}^3$ metric takes the following form 
    \begin{equation}
    \label{GRX_Metric}
    g_{\mu\nu} =  \left(
                  \begin{matrix}
                    -\frac{e^{\frac{1}{2}Q}}{\sqrt{t}} & 0 & 0 & 0 \\
                    0 & \frac{1}{2}\left(te^{-P} + A^2\right)  & \frac{1}{2}\left(te^{-P} - A^2\right) & 0 \\
                    0 & \frac{1}{2}\left(te^{-P} - A^2\right) & \frac{1}{2}\left(te^{-P} + A^2\right) & 0 \\
                    0 & 0 & 0 & t e^{P(t,r)} 
                  \end{matrix}
                  \right),
    \end{equation}  
    where  $A(t,r) = \left(\frac{e^{Q}}{t}\right)^{1/4}$. The functions $P$ and $Q$ are chosen as in the original Gowdy case, eq. (\ref{PandQ_Gowdy}), with the adequate variable substitution given by the coordinate transformation.
    
    \end{enumerate}

    \subsection{Numerical evaluation}
    
    given any spacetime metric $g_{ab}$, we can use the $3+1$ decomposition to compute the spatial metric $\gamma_{ab}$ and the extrinsic curvature $K_{ab}$ in order to evaluate the constraint equations (\ref{hamiltonian_error}) and (\ref{momentun_error}). 
    This will allow us to develop the main test that the numerical obtained solutions must pass (the fulfillment of the constraint equations). 
    Every numerical experiment in this section will be performed in a square grid, meaning $N_x = N_y = N$.
    Unless other thing stated, the parameter $t$ for Gowdy and GRX metrics take the values $t = 0.1$ and $t=0.5$ respectively;  $\phi_0=10^{-8}$ for the PFLRW, and the parameters $t$, $A$ and $d$ for MXY are $0.1$, $0.25$ and $\sqrt{2}$ respectively.\\
    
    %Here might the error be dominated by the evaluation of R? (second order derivatives on the metric \gamma)
    \begin{figure}[ht]      
    \centering 
      \includegraphics[width=0.55\textwidth,center]{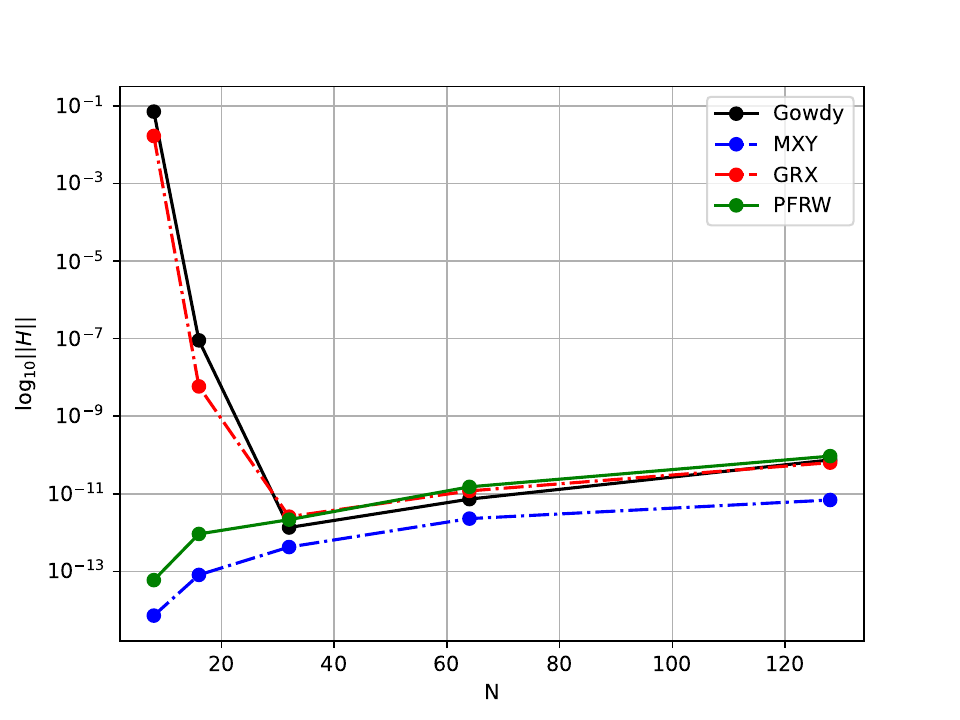}
      \caption{Hamiltonian constraint evaluation depending on the number of nodes considered in a squared grid.
    Gowdy eq. (\ref{Gowdy_Metric}), Minkowski(MXY) eq. (\ref{MXY_Metric}), Gowdy gauge wave (GRX) eq. (\ref{GRX_Metric}) and PFLRW Eq. (\ref{SPFLRW_Metric}).
      }
      \label{fig:HamiltonianError_Checker}  
    \end{figure}
    
    \begin{figure}[h!]
     \begin{subfigure}{0.5 \textwidth}
         \includegraphics[width=1.05\textwidth,left]{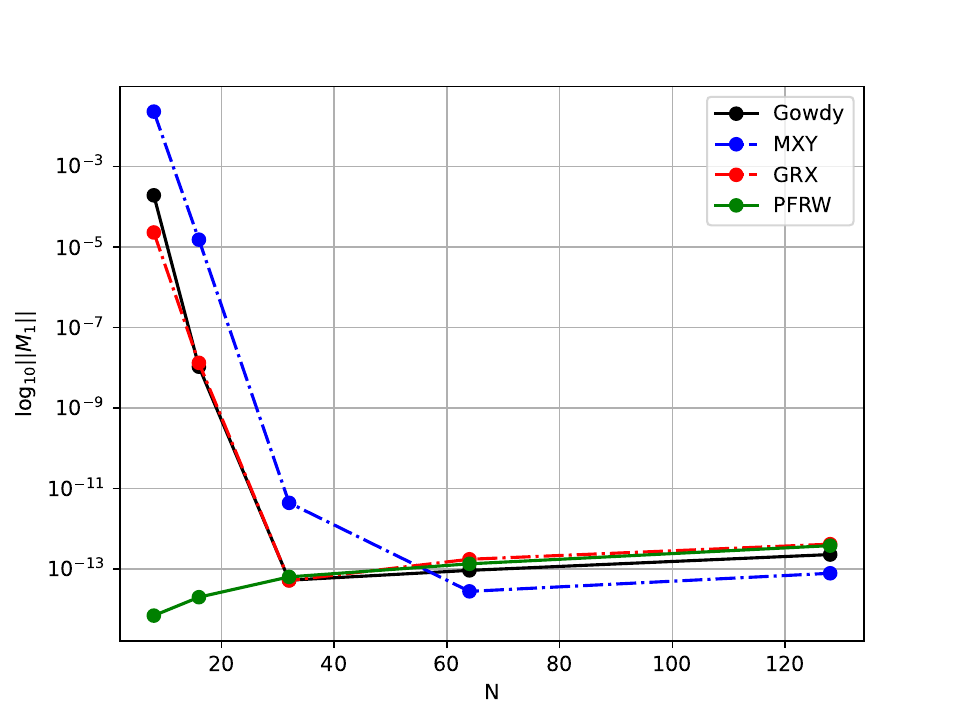}
         \caption{$M_1$}
         \label{fig:ErrorM1}
     \end{subfigure}
     \hfill
     \begin{subfigure}{0.5 \textwidth}
         \includegraphics[width=1.05\textwidth,right]{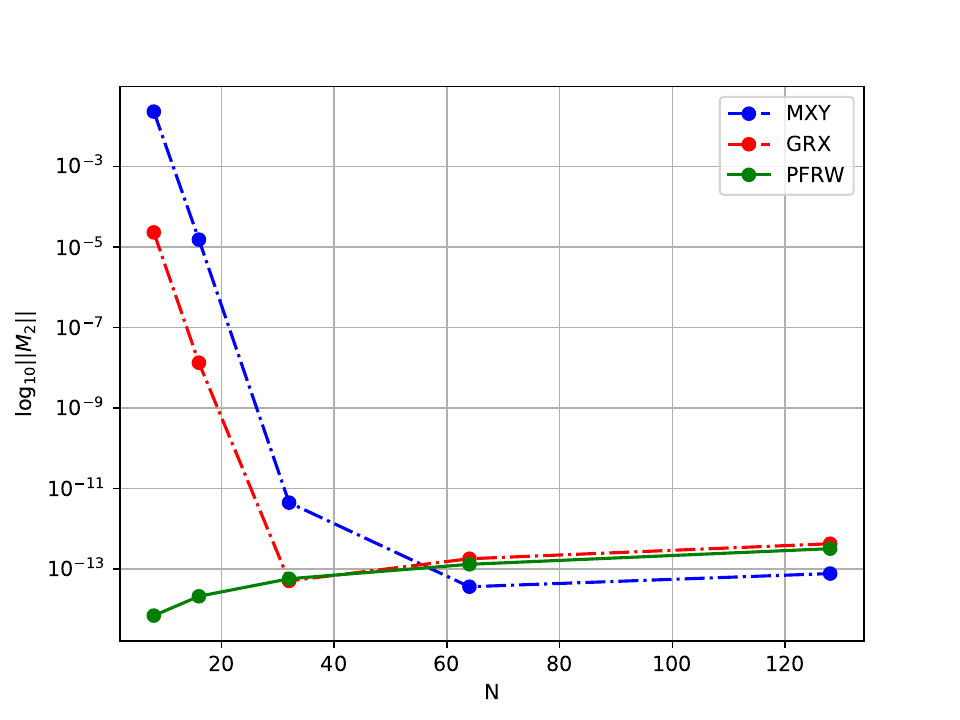}
         \caption{$M_2$}
         \label{fig:ErrorM2}
     \end{subfigure}
    \caption{First (left) and second (right) components of the Momentum constraint evaluation depending on the number of nodes of the grid.
    Gowdy eq. (\ref{Gowdy_Metric}), Minkowski(MXY) eq. (\ref{MXY_Metric}), Gowdy gauge wave (GRX) eq. (\ref{GRX_Metric}) and PFLRW eq. (\ref{SPFLRW_Metric}).}
    \label{fig:MomentumError_Checker}
    \end{figure}  
    
    In \figref{fig:HamiltonianError_Checker} we can see how the error of the evaluation of (\ref{hamiltonian_error}) changes with the number of grid nodes. Here we  observe that the difference between the behavior of the two Gowdy cases compared with Minkowski and PFLRW metrics is quite remarkable. The best error for MXY and PFLRW is obtained with the lower $N$ value. 
    In contrast to the above, when it comes to consider the Gowdy cases, a low number of nodes ($8$ or $16$) is insufficient to achieve an adequate evaluation. To understand this difference, we notice that during the evaluation of the Hamiltonian constraint, the Fourier differentiation is only used to compute the Ricci scalar $R$. Thus, this contrast is due to the complexity of the functions involved in the metrics. On the one hand, the Minkowski and PFLRW metrics depend linearly on $sine$ and $cosine$ functions, which are well resolved in small grids. On the other hand, the Gowdy metrics involve exponential functions. These functions require more nodes in order to achieve a ``fair'' representation. 
    In particular, for $N_{r}\geq32$ the behavior of the four cases is qualitatively the same with violations of the Hamiltonian constraint around $10^{-10}$ or smaller.
    The Momentum constraint, plotted in \figref{fig:MomentumError_Checker}, behaves similarly but with even smaller violations once the number of nodes is enough to resolve the derivatives of $K_{ab}$.\\

    It is work noticing that the Hamiltonian constraint for MXY and PFLRW cases in \figref{fig:HamiltonianError_Checker}, as well as the Momentum constraint for PFLRW case in \figref{fig:MomentumError_Checker}, \tit{grows} with the number of nodes. 
    In our implementation, the Fourier differentiation scheme in double precision presents an error in the order of $10^{-12}$, so it is expected that the evaluation of a highly non-linear function as the Hamiltonian or Momentum constraints is dominated by round-off errors when representing relatively simple functions. 
    Additionally, this is also due to the error on the evaluation of the Ricci curvature.
    Since it technically involves second order derivatives of the metric and we are computing them by composition of first order discrete Fourier differentiation matrices, this involves an error that can dominate for big mesh sizes.
    Although this, is worrisome for very accurate ADM constraint evaluations, it is not representative for the results presented in this work where the pathological behaviors presented in Section~\ref{section:test_error} and studied in Section~\ref{sec:Stability} are many orders of magnitude above this contribution. 

\section{Aliasing error and filtering strategy}
\label{appendix:aliasing_and_filter}

Aliasing error emerges from the non-linearities of the PDE and is due to the finiteness of the grid  (see \cite{Kopriva,Canuto}). Since our equations (eqs. (\ref{drXeq}) to (\ref{Zeq})) are non-linear, determine the relevance of this error is important to obtain accurate numerical results.\\

To illustrate this phenomena let us consider a $1-$dimensional region $[-L,L]$ and a $N$ nodes regular grid $x_i = -L + i h$ with $h = 2L/N$. Now, assume that we want to compute the product of two periodic fields  $u$ and $v$  in $[-L,L]$. Since they are periodic functions, the values of the fields $u_i$ and $v_j$ at the node $x_i$ can be decomposed in truncate Fourier series as
  \begin{align*}
    u_i  = \frac{1}{2L}\sum_{k = -N/2+1}^{N/2} \tilde{u}_k\, b_k(x_i), \quad  
    v_i  = \frac{1}{2L}\sum_{l = -N/2+1}^{N/2} \tilde{v}_l\, b_l(x_i),
  \end{align*}
where $b_k(x)$ is the Fourier basis in $1$-dimension and the numbers $\tilde{u}_k$ and $\tilde{v}_l$ are the spectral coefficients of $u_i$ and $v_i$ respectively. The numbers $k$ and $l$ are known as the \textit{wave numbers} or \tit{frequencies}. 
The product between these two fields at the node $x_i$ is given by
  \begin{equation*}
    \label{uv_Product}
    \begin{split}
      u_iv_i &=  \frac{1}{(2L)^2}\sum_{k,l= -N/2+1}^{N/2}b_{k+l}(x_i)\tilde{u}_k\tilde{v}_l,
    \end{split}
  \end{equation*}
which contains terms with frequencies $k+l$ that exceed the range $[-N/2+1,N/2]$, leading to the so-called \textit{aliased frequencies}. Since these frequencies cannot be represented in the frequency grid, due to the $k$ periodicity of the basis functions $b_k(x)$, the spectral coefficients associated to them are absorbed by their periodic \tit{equivalents} inside the grid. As a result, the computed product $u_i v_i$ may deviate significantly from the true product at the node $x_i$. This is what is known as the aliasing error, and it can be quickly accumulated during the evolution of the system (\ref{drXeq})-(\ref{Zeq}) due to non-linear terms in the equations.\\

To control this error, we can modify the Fourier series of the fields by truncating them up to some frequency $k_M$. The idea is that for any $k,l \in [-k_M,k_M]$ we have that the frequencies $k+l\in [-N/2+1,N/2]$, avoiding the aliased frequencies. For instance, it can be easily proved that (see \cite{Kopriva}) by demanding  
 \begin{equation*}
 %\label{eq:23_Filter}
  k_M \leq \frac{2}{3}(N/2),
\end{equation*}
the aliasing due to products of pairs of functions is removed. This is known as the \tit{$(2/3)-$filter} and is the easier way to control the aliasing error. Equivalently, it is easy to see that with 
\begin{equation*}
 %\label{eq:12_Filter}
    k_M \leq \frac{1}{2}(N/2),    
\end{equation*}
we can remove the aliasing due to products of triplets of functions, we call this the \tit{$(1/2)-$filter}.\\

In general, a filter can be defined as a modification of the truncated Fourier series as (see \cite{Canuto})
 \begin{align*}%\label{eq:1d_fields_filter}
  u_i  = \frac{1}{2L}\sum_{k = -N/2+1}^{N/2} \sigma_k\tilde{u}_k \ b_k(x_i),
\end{align*}
where $\sigma_k$ is called the filter function. For the \tit{$(2/3)-$} and \tit{$(1/2)-$filters} $\sigma_k$ takes the form of the step function
\begin{equation*}
    \sigma_k = 
    \begin{cases}
        1 ~~ \text{ if } ~ |k|\leq k_M \\
        0 ~~ \text{ otherwise }
    \end{cases}.
\end{equation*}\\

This is why we refer to these filters as \tit{step filters}. There are many other filtering strategies in the literature; see \cite{Canuto} for a detailed discussion about this topic. In particular, in this work we will be interested in the exponential filter where the filter function is given by
  \begin{equation}
    \label{ExpFilt_Function}
    \sigma_k := \sigma(\theta_k := \frac{2 \pi k}{N}) = e^{-\alpha \theta_k^p}, ~~~ \alpha>0 \text{ and } p\in2\mbb{N}.
  \end{equation}

\begin{figure}[ht]
\begin{subfigure}[b]{0.5\textwidth}
    \includegraphics[width=.9\textwidth,left]{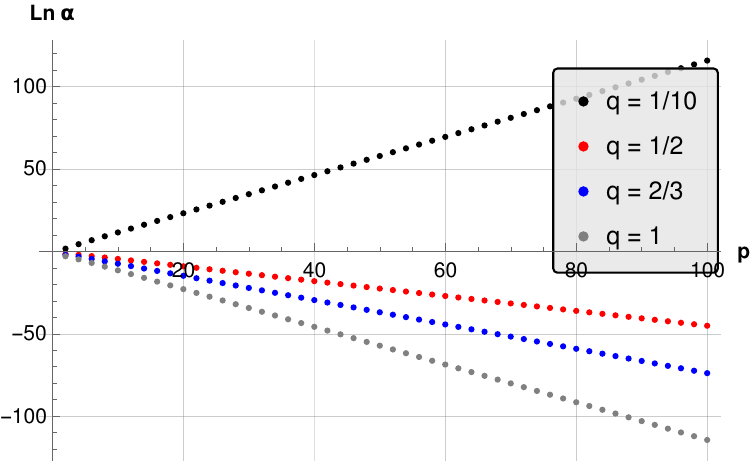}
    \caption{~}
\end{subfigure}
\hfill
\begin{subfigure}[b]{0.5\textwidth}
    \includegraphics[width=0.9\textwidth,left]{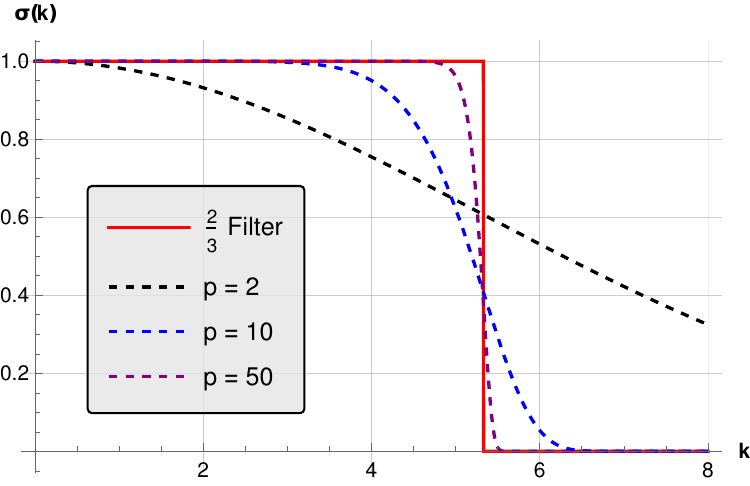}
    \caption{~}
\end{subfigure}
\caption{\tbf{(a)} Values of the parameters $p$ and $\alpha$ that satisfies that the maximum of the derivative of $\sigma(k)$ is at $k = q(N/2)$. For each $q$ there is not only a point in the parameter space that satisfies this condition, but a curve in the plane $p-\ln\alpha$. \tbf{(b)} Results for $q = 2/3$ compared with the $(2/3)-$filter. Increase the value of $p$ results in a higher value of the derivative of $\sigma(k)$ producing a faster screening of the spectral coefficients beyond $k = q(N/2)$.}
\label{Filter_q_plots}
\end{figure}

It is possible to choose combinations of the parameters $\alpha$ and $p$ that produce approximations of the step filters. To establish these combinations, notice that the first derivative of $\sigma(k)$, denoted by $\sigma'$, is zero in $k = 0$ and tends to zero when $k$ tends to infinity. This means that $\sigma'$ reaches its maximum at some $k$ in $(0,N/2]$. In the case of the step filters, the derivative is zero everywhere except at $k = q (N/2)$ (with $q = 1/2 \text{ and } q = 2/3$ we recover the \tit{$(1/2)-$filter} and the \tit{$(2/3)-$filter} respectively) where it diverges.
By identifying the exponential filter derivative's maximum with $k = q (N/2)$, a $q-$dependent relation between $\alpha$ and $p$ emerges. Therefore, we can approximate the step filters and, in a certain way, generalize them. In \figref{Filter_q_plots} we can see which combinations of parameters are related with each value of $q$.

Values of $p$ between $p=2$ and $p = 100$ can provide good approximations to step filters. From \figref{Filter_q_plots}, we see that values of $\ln\alpha \in [-100,100]$ are enough to reproduce step filters from $q = 1$ to $q=1/10$.
However, to allow more drastic filters, even those close to $q = 0$, we consider $\ln\alpha \in [-100,350]$. By exploring the space of parameters $\ln \alpha-p$, we can find the combination of $\alpha$ and $p$ that allows us the best aliasing error reduction during the radial evolution.

\subsection{Exploration of the aliasing error}\label{appendix:Aliasing}

\begin{figure}[h] 
\begin{subfigure}[b]{0.52\textwidth}
     \includegraphics[width=1.1\textwidth,center]{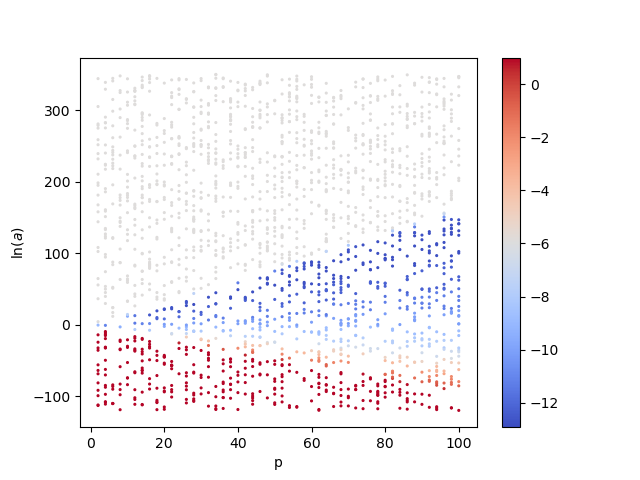}
     \caption{$N=32$}
\end{subfigure}
\begin{subfigure}[b]{0.52\textwidth}
    \includegraphics[width=1.1\textwidth,center]{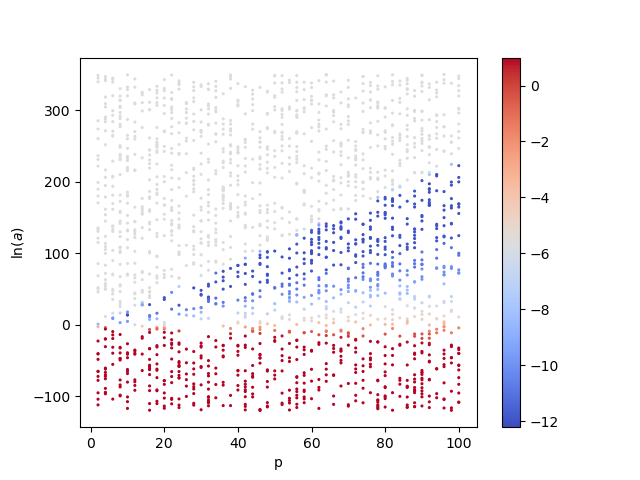}
    \caption{$N=64$}
\end{subfigure}
\caption{Result of the filtering parameter exploration for the PFLRW metric in square grids with $Factor = 32$. The color bar represents the maximum constraint joint (Hamiltonian and Momentum) constraint violation in a $\log_{10}$ scale.}
\label{FilterSPE_PFLRW}
\end{figure}

The exponential filter, defined in eq. (\ref{ExpFilt_Function}), can be used as a generalization of the step filters. It provides us control on the screening of the spectral coefficients by varying the parameters $\alpha \text{ and } p$. 
By exploring the parameter space $\ln{\alpha}-p$, we can obtain couples of values of $\alpha$ and $p$ such that the aliasing error is almost eliminated. We call this \tit{parameter filter exploration}.
When this process is applied to the PFLRW metric (see \figref{FilterSPE_PFLRW}), it allows us to obtain better constraint violations for \tit{big} grids that the obtained with the step filters (see fig. \ref{fig:PFLRWTest}\tbf{(b)}).
However, applying any of this filters do not make the method convergent.

\end{document}